\pdfoutput=1

\documentclass[11pt,a4paper]{article}

\usepackage[utf8]{inputenc}
\usepackage[T1]{fontenc}
\usepackage[english]{babel}

\usepackage[osf,sc]{mathpazo}          
\usepackage[scaled=0.88]{helvet}        
\usepackage{microtype}                  
\usepackage{ellipsis}                   

\usepackage[
    a4paper,
    top=3.2cm,
    bottom=3.2cm,
    left=3.8cm,
    right=3.8cm,
    headheight=14pt,
    headsep=1.2cm,
    footskip=1.5cm,
]{geometry}

\usepackage{setspace}
\usepackage[dvipsnames,svgnames]{xcolor}
\definecolor{ink}{HTML}{1a1a1a}
\definecolor{inksoft}{HTML}{4a4a4a}
\definecolor{accent}{HTML}{8B3A2A}       
\definecolor{linkblue}{HTML}{2B4C7E}     
\definecolor{codebg}{HTML}{F5F3EF}
\definecolor{rulegray}{HTML}{BFBAB0}

\usepackage{fancyhdr}
\fancypagestyle{firstpage}{%
    \fancyhf{}
    
    \fancyfoot[C]{\small\color{rulegray}\thepage}
}

\usepackage{titlesec}

\titleformat{\section}
    {\large\bfseries\color{ink}}
    {\thesection.}
    {0.6em}
    {}
    [\vspace{0.2em}{\color{rulegray}\hrule height 0.4pt}\vspace{0.3em}]

\titlespacing*{\section}{0pt}{2.2em}{0.8em}

\titleformat{\subsection}
    {\normalsize\bfseries\color{ink}}
    {\thesubsection}
    {0.6em}
    {}

\titlespacing*{\subsection}{0pt}{1.6em}{0.5em}

\titleformat{\subsubsection}
    {\normalsize\itshape\color{inksoft}}
    {\thesubsubsection}
    {0.6em}
    {}

\titlespacing*{\subsubsection}{0pt}{1.2em}{0.4em}

\usepackage{enumitem}
\setlist{
    topsep=0.5em,
    itemsep=0.2em,
    parsep=0.1em,
    leftmargin=1.5em,
}
\setlist[itemize]{label={\color{accent}\textbullet}}
\setlist[enumerate]{label={\color{accent}\arabic*.}}

\usepackage{graphicx}
\usepackage{float}
\usepackage{booktabs}                   
\usepackage{caption}
\usepackage{subcaption}

\usepackage{amsmath,amssymb,amsthm}

\theoremstyle{definition}

\usepackage{listings}
\usepackage[
    colorlinks=true,
    linkcolor=linkblue,
    citecolor=accent,
    urlcolor=linkblue,
    bookmarks=true,
    bookmarksnumbered=true,
    pdfstartview=FitH,
]{hyperref}

\usepackage[numbers,sort&compress]{natbib}
\renewenvironment{abstract}{%
    \vspace{0.5em}
    \noindent\begin{minipage}{\textwidth}
    \small
    \setstretch{1.1}
    \noindent{\bfseries\color{ink} Abstract.}\hspace{0.4em}\ignorespaces
}{%
    \end{minipage}
    \vspace{1em}
}

\newcommand{\keywords}[1]{%
    \vspace{0.3em}
    \noindent{\small\textbf{Keywords:}\hspace{0.4em}\textit{#1}}
    \vspace{0.5em}
}

\usepackage{tcolorbox}
\tcbuselibrary{breakable,skins}

\newtcolorbox{callout}[1][]{%
    enhanced,
    breakable,
    colback=codebg,
    colframe=codebg,
    boxrule=0pt,
    borderline west={2.5pt}{0pt}{accent},
    left=10pt,
    right=10pt,
    top=8pt,
    bottom=8pt,
    fontupper=\small,
    before upper={\textbf{\color{accent}\small #1}\par\vspace{0.3em}},
}

\newcommand{\shorttitle}{Short Title Here}

\newcommand{\articletitle}{PRISMA: Toward a Normative Information Infrastructure\\[0.15em] for Responsible Pharmaceutical Knowledge Management}

\begin{document}
\thispagestyle{firstpage}
\renewcommand{\shorttitle}{PATOS--Lector--PRISMA Architecture}

{
\centering

\vspace*{0.5cm}

{\Large\bfseries\color{ink}\articletitle\par}

\vspace{1.5em}

{\normalsize
\textbf{Eugenio Rodrigo Zimmer Neves}\footnotemark,\quad
\textbf{Amanda Vanon Correa},\quad
\textbf{Camila Campioni},\quad
\textbf{Gabielli Pare Guglielmi},\quad
\textbf{Bruno Morelli}
}

\vspace{0.5em}

{\small\color{inksoft}
Pharmdata, Porto Alegre, Brazil
}

\vspace{0.3em}

\footnotetext{Corresponding author: \href{mailto:eugenio@pharmdata.com.br}{eugenio@pharmdata.com.br}}

{\small\color{rulegray} February 2026}

\vspace{0.3em}

{\color{rulegray}\hrule height 0.4pt}

\par
}

\vspace{0.5em}

\begin{abstract}
Most existing approaches to AI in pharmacy collapse three epistemologically distinct operations into a single technical layer: document preservation, semantic interpretation, and contextual presentation. This conflation is a root cause of recurring fragilities including loss of provenance, interpretive opacity, alert fatigue, and erosion of accountability. This paper proposes the PATOS--Lector--PRISMA (PLP) infrastructure as a normative information architecture for responsible pharmaceutical knowledge management. PATOS preserves regulatory documents with explicit versioning and provenance; Lector implements machine-assisted reading with human curation, producing typed assertions anchored to primary sources; PRISMA delivers contextual presentation through the RPDA framework (Regulatory, Prescription, Dispensing, Administration), refracting the same informational core into distinct professional views. The architecture introduces the Evidence Pack as a formal unit of accountable assertion (versioned, traceable, epistemically bounded, and curatorially validated), with assertions typified by illocutionary force. A worked example traces dipyrone monohydrate across all three layers using real system data. Developed and validated in Brazil's regulatory context, the architecture is grounded in an operational implementation comprising over 16,000 official documents and 38 curated Evidence Packs spanning five reference medications. The proposal is demonstrated as complementary to operational decision support systems, providing infrastructural conditions that current systems lack: documentary anchoring, interpretive transparency, and institutional accountability.
\end{abstract}


{\color{rulegray}\hrule height 0.4pt}
\vspace{1em}


\section{Introduction}

The incorporation of artificial intelligence into pharmacy practice has been advancing at an accelerating pace. Comprehensive reviews document growing applications in drug interaction detection, prescription validation, dispensing automation, adverse event prediction, and therapeutic reconciliation support \cite{Chalasani2023,MottaghiDastjerdi2024,Hamishehkar2025}. Operational systems such as NoHarm.ai demonstrate measurable clinical impact in real hospital environments, with substantial reductions in prescription error rates and significant expansion of pharmaceutical review coverage \cite{DosSantos2019,Leitao2023}. The emergence of large language models has introduced an additional dimension to the field, with evaluations indicating performance comparable to that of human pharmacists in specific medication review tasks \cite{Roosan2024,Shin2024}, albeit with significant degradation as clinical complexity increases \cite{Blotske2025}.

However, this technological advance coexists with a persistent structural fragility. The literature on pharmacy information systems provides empirical evidence that the introduction of technology upon misaligned informational layers can amplify latent failures and generate unintended consequences: documentary fragmentation, cognitive overload, task duplication, and progressive degradation of information quality \cite{Rohani2023,Harrison2007,Campbell2006}. The problem, therefore, is not the absence of computational tools, but the absence of an \emph{infrastructure} capable of coherently organizing the different strata of responsibility involved in the production, interpretation, and presentation of pharmaceutical knowledge. This diagnosis is not exclusive to the pharmaceutical domain: recent analyses of the broader enterprise AI ecosystem converge on the same conclusion. The MIT NANDA (Networked Agents And Decentralized Architecture) report, based on a systematic review of 300 public AI initiatives and interviews with 52 organizations, demonstrates that despite \$30--40 billion in enterprise generative AI investments, 95\% of organizations achieve zero return and only 5\% of custom pilots reach production \cite{ChallapallyMITNANDA2025}. Failures are attributed to fragile workflows, lack of contextual learning, and operational misalignment. Particularly relevant to this work, the \emph{Healthcare \& Pharma} sector receives one of the lowest disruption scores among nine analyzed sectors (0.5 out of 5.0), with signals limited to documentation and transcription pilots and no alteration of underlying clinical models. Ramesh \cite{Ramesh2026}, commenting on these findings, argues that weak foundations are not the root problem but rather a symptom of how organizations make decisions, reward work, and learn under uncertainty.

The root of this misalignment can be identified in a precise conceptual gap: most existing approaches (whether clinical decision support systems, Retrieval-Augmented Generation (RAG) pipelines, conversational agents, or rule engines) operate on already-transformed data, without preserving the formal distinction between \emph{primary source}, \emph{interpretation}, and \emph{recommendation}. The original regulatory source (package insert, monograph, clinical guideline, normative act) is typically ingested, fragmented, and reprocessed until its provenance becomes irrecoverable. Talisman \cite{Talisman2026b} articulates this mechanism precisely: during training, billions of documents are compressed, merged, and statistically distributed across billions of parameters in a process that is \emph{lossy} and irreversible, structurally destroying provenance. This destruction operates on two layers: retrieval, which can cite sources, and parametric, where provenance is irrecoverably lost. The result is the phenomenon of \emph{vibe citing}, in which fabricated references with legitimacy markers contaminate subsequent knowledge bases. Interpretation, a cognitive act requiring context, judgment, and professional responsibility, is frequently delegated to statistical models without explicit traceability mechanisms. And the presentation to the end user regularly conflates system output with source authority, eroding the boundaries of responsibility between machine, professional, and institution.

Recent works reinforce, independently and without coordination, that this fragility is recognized by the scientific community. Sabel et al.\ demonstrate that even in a highly controlled scenario (counseling on a single over-the-counter medication), the application of large language models requires the progressive introduction of normative layers, control YAMLs, decision graphs, specialized agents, and explicit human supervision \cite{Sabel2025}. The probabilistic model, by itself, does not offer the accountability guarantees demanded by the domain. This empirical finding functions as exogenous validation of a broader structural problem: the central challenge of pharmaceutical information is not algorithmic but architectural.

It is precisely in this space that this paper proposes the \textbf{PATOS--Lector--PRISMA} set as a normative information infrastructure for responsible pharmaceutical knowledge management. The proposal articulates three functionally distinct and conceptually complementary layers. \textbf{(i) PATOS} is a sovereign document preservation layer oriented toward explicit versioning and primary evidence integrity, grounded in the principles of the OAIS (Open Archival Information System) model and trusted digital repositories \cite{CCSDS2012,OCLCCRL2007}. \textbf{(ii) Lector} is a machine-assisted reading and technical curation environment, conceived as a space of cognitive mediation between the preserved source and professional interpretation, drawing on the tradition of \emph{computational sensemaking} \cite{PirolliRussell2011} and \emph{human-in-the-loop systems} \cite{Kittur2013}. \textbf{(iii) PRISMA} is a contextual presentation layer that organizes knowledge into versioned assertions accompanied by formal evidence trails (\emph{Evidence Packs}), regulatory context graphs, and explicit argumentative structures, grounded in the literature on data provenance \cite{Moreau2013,Buneman2001}, knowledge graphs \cite{Hogan2021}, and computational argumentation \cite{Toulmin1958}.

The design of this infrastructure proceeds from a deliberate principle: the system does not decide, does not prescribe, and does not authorize. It preserves sources, organizes the interpretive space, and documents the knowledge construction trajectory, keeping responsibility firmly in the human and institutional domain. This position finds direct grounding in the sociotechnical systems literature \cite{Suchman1987}, in evidence on health automation risks \cite{Harrison2007,Ash2007}, and in contemporary studies on artificial intelligence governance \cite{Rahwan2019,Floridi2018}.

This paper offers four main contributions:

\begin{enumerate}
    \item \textbf{Formulation of the architectural problem}: identifies and characterizes the structural gap between preservation, interpretation, and presentation in pharmaceutical knowledge management, arguing that this gap is the root cause of recurring fragilities in health information systems.

    \item \textbf{Proposal of the PATOS--Lector--PRISMA infrastructure}: presents a three-layer architecture, conceptually grounded in established academic traditions, oriented toward traceability, auditability, and explicit delineation of responsibilities.

    \item \textbf{Introduction of the \emph{Evidence Pack} concept}: proposes a formal evidence packaging unit (versioned, traceable, and contextualized) as the central mechanism for sustaining assertions in regulatory and clinical domains.

    \item \textbf{Typification of pharmaceutical assertions by illocutionary force}: classifies assertions according to their normative nature (indication, contraindication, warning, precaution, normative silence, among others), grounded in speech act theory and deontic logic, including the formal treatment of normative silence as an epistemically relevant category.
\end{enumerate}

The paper is structured as follows. Section~2 positions the proposal in relation to existing literature, articulating the theoretical foundations that underpin each layer. Section~3 describes the proposed architecture and its design principles. Section~4 presents NoHarm.ai as a motivating case, demonstrating how the infrastructure would complement, without replacing, operational decision support systems. Section~5 discusses architectural, regulatory, and institutional implications. Section~6 concludes with a synthesis of contributions and future directions.

\section{Related Work}

The architecture proposed by the PATOS / Lector / PRISMA set is situated at the convergence of multiple established academic traditions that, while mature in their respective domains, remain historically fragmented when applied to pharmaceutical and regulatory knowledge management. These domains include digital document preservation, machine-assisted reading, evidence provenance and traceability, contextual knowledge representation, explainable artificial intelligence, and sociotechnical studies of health information systems. The originality of the proposal lies not in an incremental advance within a single subfield, but in the coherent architectural integration of these foundations, explicitly oriented toward regulatory, clinical, and institutional accountability.

\subsection{Document preservation, versioning, and regulatory evidence sovereignty}

The conceptual separation between sovereign document preservation layers and interpretive or analytical layers finds direct grounding in the \emph{Open Archival Information System} (OAIS) model, formalized as ISO 14721, which defines structural principles for the ingestion, storage, management, and access of preserved digital objects, with emphasis on authenticity, integrity, explicit versioning, and independence from secondary use \cite{CCSDS2012}. Complementarily, the \emph{Trusted Digital Repositories} model reinforces that the reliability of a repository derives not from its applied purpose, but from the verifiable robustness of its control, audit, and institutional governance mechanisms \cite{OCLCCRL2007}.

In the context of pharmaceutical and regulatory information, these principles assume critical relevance, since documents such as package inserts, monographs, clinical guidelines, and normative acts possess temporal validity, successive versions, and direct legal effects. Works in digital document engineering, such as those by Lagoze et al., anticipate this need by conceiving \emph{digital objects} as entities rich in metadata, relationships, and transformation history, transcending the view of documents as passive files \cite{Lagoze2006}. These foundations underpin the role of PATOS as a normative documentary layer, whose primary function is to preserve sources with sovereignty, explicit versioning, and traceability, without anticipating clinical or regulatory interpretations.

\subsection{Machine-assisted reading, curation, and computational sensemaking}

The conception of Lector follows the classical literature on \emph{computational sensemaking}, particularly the models of Pirolli and Russell, which describe the analysis of complex information as an iterative process in which computational systems assist the organization, filtering, and externalization of human reasoning, without replacing the analyst's judgment \cite{PirolliRussell2011}. This approach is widely adopted in domains where interpretation depends on context, exceptions, and ambiguity, all central to pharmaceutical and regulatory information.

Studies on \emph{exploratory search} reinforce this perspective by demonstrating that interaction with large documentary corpora is an open cognitive process, guided by provisional hypotheses, successive reformulations, and incremental construction of understanding \cite{Hearst2009}. In the specific field of technical and scientific reading, Marshall demonstrates that digital reading is intrinsically active, annotative, and interpretive, even in computational environments, invalidating models that treat documents merely as inputs for automatic answers \cite{Marshall1997,Marshall2009}.

Convergently, the literature on \emph{human-in-the-loop systems} establishes that the explicit incorporation of the human in the interpretive cycle is not a technological limitation but a structural requirement for systems operating in sensitive domains where interpretation errors can generate clinical, legal, or institutional consequences \cite{Kittur2013}. These works ground Lector as an environment for assisted reading, technical curation, and cognitive mediation, rather than as an autonomous agent of normative interpretation.

\subsection{Provenance, traceability, and explainability in decision support systems}

The requirement for explicit evidence traceability presented by PRISMA finds solid grounding in the \emph{data provenance} literature. The PROV model, standardized by the W3C, establishes a formal framework for representing entities, activities, and agents involved in the production, transformation, and use of data, enabling the reconstruction of origin and responsibility chains \cite{Moreau2013}. Classical works by Buneman, Khanna, and Tan deepen this notion by distinguishing \emph{where-provenance} and \emph{why-provenance}, emphasizing that the explanation of a result depends both on its origin and on the conditions that justify its existence \cite{Buneman2001}.

These concepts converge directly with the contemporary literature on \emph{Explainable Artificial Intelligence} (XAI), in which explainability is treated as a systemic and contextual property, rather than as an optional attribute of isolated statistical models \cite{DoshiVelezKim2017}. In the health domain, this requirement is reinforced by emerging regulations and by studies demonstrating that opaque decision support systems can introduce significant clinical and legal risks.

The urgency of this requirement is amplified by the finding that large language models destroy provenance in a structurally irreversible manner. Talisman \cite{Talisman2026b} demonstrates that LLM training operates as a lossy compression process that dissolves the documentary chains connecting knowledge to its origins, rendering provenance irrecoverable at the parametric level. The author traces the formalization of provenance as an archival principle from 1841 to Berners-Lee's linked data systems, observing that decades of research in computational provenance have been ignored by the generative AI industry. In the pharmaceutical domain, this phenomenon is particularly grave: a fabricated drug interaction or a hallucinated contraindication is not an academic inconvenience but a direct risk to patient safety. In this context, the concept of \emph{Evidence Pack}, although originally formulated within the scope of PRISMA, aligns conceptually with the requirement that every presented assertion be accompanied by traceable, contextualized, and auditable evidence, and responds directly to the challenge of preserving provenance as an intrinsic architectural property, not as metadata added \emph{a posteriori}.

\subsection{Knowledge graphs, context, and regulatory argumentation}

The literature on \emph{knowledge graphs} provides the structural foundation for PRISMA's presentation layer, conceiving knowledge as a network of semantically typed entities and relationships \cite{Hogan2021}. In the pharmaceutical domain, this approach has been explored to represent relationships between substances, indications, contraindications, specific populations, and clinical evidence, though frequently dissociated from documentary provenance.

Complementarily, the \emph{contextual graphs} proposed by Br\'{e}zillon formalize context as an explicit entity, enabling the representation of validity variations associated with specific conditions of use, clinical scenario, or regulatory framing \cite{Brezillon2005}. In the field of computational argumentation, models inspired by Toulmin offer a framework for structuring claims, evidence, qualifiers, and rebuttals, which is particularly relevant in regulatory domains where the strength of an assertion depends on its explicit conditions of support \cite{Toulmin1958}. These foundations underpin the conception of PRISMA as a contextual presentation layer of grounded assertions, rather than as an automatic response system.

Convergently, the contemporary distinction between \emph{semantic layers} (metric governance for human consumption) and \emph{context graphs} (decision infrastructure with traceability and reasoning) precisely articulates the challenge that the architecture proposed here addresses: AI systems do not need layers that standardize calculations, but formal representations that support reasoning about concepts, relationships, and domain inference rules \cite{Talisman2026}. Talisman observes that the industry has invested extensively in semantic layers to make metrics consistent, but that ``meaning isn't the same as measurement'': metrics do not explain causality nor support reasoning. This observation, formulated in the context of enterprise data engineering, maps directly to the architecture proposed here: PRISMA is not a pharmaceutical \emph{semantic layer} (a standardized metric presentation layer), but a contextual presentation layer sustained by sovereign document preservation (PATOS) and qualified interpretation with human curation (Lector).

\subsection{Sociotechnical systems, systemic risks, and governance}

The proposal dialogues directly with the sociotechnical systems literature, which rejects the notion of fully deterministic decisions in complex environments. Suchman demonstrates that human action is always situated, contextual, and partially unpredictable, rendering the full delegation of authority to automated systems inadequate \cite{Suchman1987}. In the health context, classical studies demonstrate that information technologies frequently introduce unforeseen consequences arising from the interaction between systems, professionals, processes, and organizations \cite{Harrison2007,Ash2007,Campbell2006}.

More recent works on machine behavior and AI governance reinforce the need to understand intelligent systems as institutional components whose effects extend beyond the technical domain \cite{Rahwan2019,Floridi2018}. This perspective explicitly grounds the position adopted by PRISMA, according to which the system structures knowledge and documents the interpretive trajectory rather than assuming decisional authority, preserving human and institutional responsibility.

\subsection{Epistemological foundations: information, interpretation, and presentation}

The architectural separation between preservation, interpretation, and presentation proposed in this work is neither arbitrary nor merely technical; it finds convergent grounding in established epistemological traditions in information science, philosophical hermeneutics, and the philosophy of information.

Buckland, in foundational work, distinguishes three senses of ``information'' that are frequently, and detrimentally, conflated: \emph{information-as-thing} (the document, the datum, the tangible artifact), \emph{information-as-process} (the cognitive act through which one becomes informed), and \emph{information-as-knowledge} (the resulting understanding, that which is perceived and comprehended through the process) \cite{Buckland1991}. This tripartite distinction maps directly to the proposed architecture: PATOS preserves information-as-thing (the regulatory document as artifact); Lector operates on information-as-process (technical reading, assisted interpretation); and PRISMA presents information-as-knowledge (contextualized understanding, organized for professional decision-making). Buckland's contribution is relevant less as analogy than as admonition: systems that conflate these three senses tend to treat the document as if it were the knowledge, or the knowledge as if it were the document, both distinct forms of informational failure.

Convergently, the hermeneutic tradition, as applied to information science by Capurro and Hjørland \cite{Capurro2003} and to information systems by Myers \cite{Myers1995}, provides support for the thesis that the separation between preserved text, interpretation, and application is not a technical convenience but an epistemological necessity. The hermeneutic arc described by Ricoeur articulates three mutually constitutive moments: \emph{explanation} (structural analysis of the text as an autonomous object), \emph{understanding} (discovery of meaning through mediation between the world of the text and the world of the reader), and \emph{appropriation} (assimilation of interpretation into the reader's horizon of action) \cite{Ricoeur1976,Myers1995}. Ricoeur insists that the text, once fixed by writing, acquires autonomy from the author and the original context of production, becoming an interpretable object in its own right. This position directly grounds PATOS's principle of documentary sovereignty: the preserved document is not a representation of its author's thought, but an autonomous artifact that can and should be read, re-read, and interpreted under different contextual conditions, without its integrity being compromised.

In contemporary philosophy of information, Floridi proposes the \emph{Levels of Abstraction} (LoA) method, according to which any system can be legitimately analyzed at different levels, each defined by its own set of observables and operations \cite{Floridi2011}. Different LoAs of the same system are not competing descriptions but complementary perspectives, each suitable for a specific type of analysis or action. This formalization provides the most rigorous philosophical justification for the multi-layer design: PATOS, Lector, and PRISMA are neither redundant nor competing, but operate at genuinely distinct levels of abstraction over the same pharmaceutical informational reality, each with its own observables, legitimate operations, and validity criteria.

\subsection{Normative typification, speech acts, and evidence qualification}

Pharmaceutical information is intrinsically normative: a package insert does not simply \emph{describe} a medication's properties; it \emph{prescribes} conduct, \emph{warns} about risks, \emph{prohibits} uses, and in certain cases, \emph{remains silent} on clinically relevant questions. This illocutionary heterogeneity (the fact that different assertions within the same document possess radically distinct communicative forces) requires specific theoretical grounding.

Speech act theory, developed by Austin and formalized by Searle, distinguishes three dimensions of every utterance: the \emph{locutionary} act (producing a meaningful expression), the \emph{illocutionary} act (what one does \emph{in saying}: asserting, warning, requesting, prohibiting), and the \emph{perlocutionary} act (the effect produced on the interlocutor) \cite{Austin1975}. Searle refines this distinction into five categories of illocutionary acts: \emph{assertives} (commit the speaker to the truth of a proposition), \emph{directives} (attempts to get the interlocutor to act), \emph{commissives} (speaker commitments), \emph{expressives} (expressions of psychological states), and \emph{declaratives} (which create states of affairs by the very act of enunciation) \cite{Searle1979}. In the context of pharmaceutical information, a therapeutic indication is a regulatory \emph{assertive}; a contraindication is a \emph{directive} (instructs the professional to avoid); a warning combines assertive and directive; and normative silence (the absence of regulatory pronouncement on a clinically relevant question) carries its own illocutionary force as non-commitment. This taxonomy grounds the assertion typification that Lector performs and that PRISMA presents in differentiated form.

Complementarily, the tradition of deontic logic applied to normative systems, as systematized by Navarro and Rodr\'{i}guez \cite{Navarro2014} from the foundational work of Alchourr\'{o}n and Bulygin \cite{AlchourronBulygin1971}, formalizes the distinction between normative and descriptive propositions, demonstrating that normative systems (composed of obligations, permissions, and prohibitions) possess a logical structure fundamentally distinct from descriptive systems. This tradition introduces the concept of \emph{normative gap}, situations in which the normative system neither permits nor prohibits an action, generating genuine indeterminacy \cite{Prakken2015}. For pharmaceutical information, this concept is directly relevant: a package insert that does not mention a drug interaction creates a normative gap; silence \emph{does not} equate to permission or safety. The formal distinction between absence of prohibition and explicit permission is precisely the kind of distinction that systems based on probabilistic models tend to collapse, and that the architecture proposed in this work seeks to preserve.

The theory of communicative action by Habermas, as applied to information systems by Ngwenyama and Lee \cite{Ngwenyama1997} and by Mingers \cite{Mingers1995}, adds an additional dimension. Habermas argues that every speech act oriented toward understanding simultaneously raises three validity claims: \emph{truth} (the proposition corresponds to the external world), \emph{normative rightness} (the act conforms to socially legitimate norms), and \emph{sincerity} (the speaker authentically expresses their internal state) \cite{Habermas1984}. Ngwenyama and Lee demonstrate that informational richness is not an intrinsic property of content but depends on the social context in which validity claims are raised and redeemed \cite{Ngwenyama1997}. A pharmaceutical monograph simultaneously claims factual truth (pharmaceutical data), normative rightness (regulatory conformity), and institutional sincerity (the manufacturer's disclosure obligation). This threefold requirement explains why the different axes of the RPDA framework (Section~3.4) need to treat the same information from distinct perspectives. Each axis prioritizes different validity claims, appropriate to the decision-making context and the professional responsibility involved.

Finally, the tradition of evidence-based medicine provides specific grounding for the \emph{Evidence Pack} concept. Sackett defines evidence-based practice as ``the conscientious, explicit, and judicious use of current best evidence in making decisions about the care of individual patients'' \cite{Sackett1996}, a formulation that inherently requires separating the evidence base (preservation), critical appraisal (interpretation), and contextualized clinical application (presentation). The GRADE (Grading of Recommendations Assessment, Development and Evaluation) framework complements this perspective by establishing that the quality of evidence and the strength of recommendation are independent dimensions: a strong recommendation can be based on moderate-quality evidence, and vice versa, which justifies why the interpretive layer must qualify, and not simply transmit, pharmaceutical information \cite{Guyatt2008}.

\subsection{Artificial intelligence in pharmacy practice}

Comprehensive reviews demonstrate that AI has been increasingly applied across multiple dimensions of pharmacy practice, including drug interaction detection, prescription validation, dispensing automation, dose optimization, and therapeutic adherence \cite{Chalasani2023,Raza2022,MottaghiDastjerdi2024}. More recent evaluations indicate direct impact on clinical and hospital pharmacy, with error reduction and improvement in operational efficiency \cite{Hamishehkar2025}.

In the hospital context, studies report applications focused on the identification of potentially inappropriate prescriptions, adverse event prediction, and high-risk patient prioritization \cite{Gosselin2021,GonzalezPerez2024}. However, systematic reviews point to high methodological bias risk and dependence on local databases, which limits the generalization of results \cite{Johns2024}. The exploration of unstructured electronic health record data through natural language processing techniques reinforces the role of AI as mediator between large volumes of clinical information and patient safety-oriented pharmacy practice \cite{DelRioBermudez2020}. Significantly, reviews on clinical decision support systems in specific contexts such as pediatrics reveal that alert-based systems exhibit \emph{override} rates between 63\% and 95\%, indicating that the interruption-and-block model, when dissociated from adequate cognitive mediation, tends to be systematically ignored by professionals \cite{StultzNahata2012}.

The emergence of large language models has introduced a new dimension to the field. Empirical evaluations indicate that these models can achieve performance comparable to human pharmacists in specific tasks such as medication review, interaction identification, and therapeutic reconciliation, especially in well-delimited scenarios \cite{Roosan2024,Shin2024,Sridharan2024}. However, benchmarking studies reveal uneven performance across models, reduced stability as clinical complexity increases, and consistent absence of explicit justification and traceability mechanisms for generated recommendations \cite{Blotske2025}. These findings reinforce that while language models substantially expand the semantic capacity of pharmacy practice support systems, they also intensify challenges related to governance, explainability, and delineation of interpretive responsibility. A systematic review of 89 studies on large language models (LLMs) in patient care confirms this picture by identifying structural limitations in both \emph{design} (data opacity, lack of optimization for the medical domain) and \emph{output} (incompleteness in 76.4\% of studies, incorrectness in 87.6\%, with hallucination and confabulation as distinct and recurrent phenomena) \cite{Busch2025}.

Despite these advances, the literature converges remarkably on a critical point: AI is widely conceived as decision support, not as a substitute for professional judgment. Empirical studies evidence predominantly positive attitudes toward AI adoption, accompanied by an explicit demand for human supervision, transparency, and institutional control \cite{Jarab2023}. Even more propositional works, which introduce concepts such as ``pharmacointelligence,'' acknowledge that the central challenge lies not only in algorithmic performance but in the organization of the informational ecosystem that sustains pharmacy practice \cite{Hatem2024}.

This finding is corroborated empirically at scale by the MIT NANDA report on the state of AI in business, which reveals what the authors term the \emph{GenAI Divide}: a division between organizations that extract real value from AI and the vast majority that remain trapped in pilots with no measurable impact \cite{ChallapallyMITNANDA2025}. The \emph{Healthcare \& Pharma} sector sits firmly on the unfavorable side of this divide, with a disruption index of 0.5 out of 5.0, the lowest alongside financial services and advanced manufacturing. The observed signals are limited to documentation and transcription pilots, with no evidence of structural change in clinical models. The report identifies that tools involving \emph{complex internal logic, opaque decision support, or optimization based on proprietary heuristics}, precisely the characteristics of clinical decision support systems, rank among the categories with the greatest adoption difficulty. This empirical finding, produced independently of the pharmaceutical domain, reinforces the central hypothesis of this work: the problem is not the absence of capable algorithms, but the absence of infrastructures that coherently organize the information upon which these algorithms operate.

\subsection{Pharmaceutical AI systems in production: the NoHarm.ai case}

Beyond the academic literature, one observes the consolidation of pharmaceutical AI systems in production environments, with measurable clinical and institutional impact. A particularly relevant example is NoHarm.ai, a Brazilian clinical decision support system focused on hospital pharmacy, widely adopted in public and private institutions. The system performs automatic real-time analysis of electronic medical prescriptions, identifying drug interactions, inadequate doses, therapeutic duplicities, allergies, route incompatibilities, and high-risk patient profiles, integrating directly with hospital systems through standards such as HL7 and FHIR.

Scientific studies associated with NoHarm.ai demonstrate substantial reductions in prescription error rates and significant increases in pharmaceutical review coverage, even without expansion of clinical teams, evidencing the potential of AI as an amplifier of the clinical pharmacist's professional capacity \cite{DosSantos2019,Leitao2023}. The initiative also stands out for its operational scale, with millions of prescriptions analyzed monthly, and for its international recognition.

From a conceptual standpoint, NoHarm.ai represents the state of the art in operational pharmaceutical decision support systems, oriented toward patient safety and clinical efficiency. However, like most solutions described in the literature, its scope focuses on real-time risk detection and mitigation, not explicitly addressing the structural organization of regulatory pharmaceutical knowledge, versioned document preservation, or the construction of formal interpretable evidence trails.

\subsection{Evidence of systemic risk in pharmacy information systems}

Beyond functional limitations, the health informatics literature provides empirical evidence that pharmacy information systems can introduce unintended systemic risks when incorporated into clinical environments without an information architecture explicitly designed for sociotechnical requirements. Rohani and Yusof identified 28 unintended consequences associated with the use of \emph{Pharmacy Information Systems}, demonstrating that factors such as system rigidity, documentary fragmentation, cognitive overload, and institutional \emph{workarounds} progressively degrade information quality and increase the risk of medication errors \cite{Rohani2023}.

These findings are reinforced by specific evidence on errors generated, not merely undetected, by electronic prescribing systems. Brown et al.\ identified eight interconnected categories of errors induced by CPOE systems, including selection errors in \emph{drop-down} menus, passive acceptance of \emph{defaults}, medication visualization fragmentation across partial screens, and \emph{workarounds} that circumvent safety mechanisms, demonstrating that failures in human-centered design directly convert into clinical risks \cite{Brown2016}. Complementarily, Feather et al.\ demonstrated that the documentation of clinical indication in electronic prescriptions, a practice recommended by health authorities to link each medication to its therapeutic justification, remains rare in practice, and that implementation attempts encounter extensive selection lists, prescriber resistance, and \emph{workarounds} by junior professionals, resulting in indication errors that degrade prescriptive decision traceability \cite{Feather2023}. This dissociation between the prescribed medication and its documented clinical justification illustrates, at the operational level, exactly the fragmentation between action and evidence that the infrastructure proposed in this paper seeks to address structurally.

These findings corroborate earlier investigations into errors induced by health information systems and reinforce that the introduction of automation upon fragile informational layers tends to amplify latent failures, transforming local limitations into systemic risks \cite{Harrison2007}. Of particular relevance for AI-based systems, these structural failures tend to propagate and amplify when new automated components are coupled to already-misaligned informational layers.

This evidence does not constitute an argument against digitization or the use of artificial intelligence in pharmacy practice, but rather points to the need for infrastructures capable of explicitly addressing information provenance, evidence traceability, and the clear delineation of responsibilities between humans and systems.

\subsection{Synthesis and contribution positioning}

Considered as a whole, the literature reveals a field that has matured rapidly in terms of point AI applications, but remains structurally lacking in normative information infrastructures that preserve documents, versions, regulatory context, and explicit evidence trails. Most existing approaches operate on already-transformed data, without guaranteeing the formal distinction between primary source, interpretation, and recommendation. The epistemological foundations reviewed, from Buckland's tripartite distinction \cite{Buckland1991} to the hermeneutic arc applied to information science \cite{Ricoeur1976,Capurro2003,Myers1995}, from speech act theory \cite{Austin1975,Searle1979} to the deontic logic of normative systems \cite{AlchourronBulygin1971,Navarro2014}, from evidence-based medicine \cite{Sackett1996,Guyatt2008} to the theory of communicative action applied to information systems \cite{Habermas1984,Ngwenyama1997}, converge in demonstrating that this separation is not a technical convenience but an epistemological and normative necessity.

It is precisely in this space that the PATOS / Lector / PRISMA proposal is positioned, treating artificial intelligence not as a substitute for regulatory or clinical authority, but as a component of an infrastructure oriented toward the responsible, traceable, and auditable interpretation of pharmaceutical knowledge, positioning itself not as a point AI application but as a structural proposal for a new generation of health information infrastructures.

\section{The PATOS--Lector--PRISMA Architecture}

\subsection{Design principles and epistemological foundations}

The PATOS--Lector--PRISMA architecture was conceived from a deliberate principle of functional separation between three cognitive and institutional operations that, in current pharmacy information systems practice, are frequently collapsed into a single technical layer: documentary \emph{preservation}, semantic \emph{interpretation}, and contextual \emph{presentation}. This separation finds convergent grounding in Buckland's tripartite distinction between information-as-thing, information-as-process, and information-as-knowledge \cite{Buckland1991}, in the hermeneutic arc between explanation, understanding, and appropriation \cite{Ricoeur1976,Myers1995}, and in Floridi's levels of abstraction method \cite{Floridi2011}. Each of these operations possesses its own requirements of governance, temporality, responsibility, and auditability, and their fusion (characteristic of most existing systems) constitutes, in the view defended here, the root cause of recurring fragilities such as loss of provenance, interpretive opacity, and erosion of the boundaries of responsibility between machine, professional, and institution.

The architecture adopts five guiding principles:

\begin{enumerate}
    \item \textbf{Documentary sovereignty}: regulatory, clinical, and normative sources are preserved in their original integrity, explicitly versioned, and protected against destructive transformations. No interpretive or presentational layer can alter, rewrite, or replace the source document.

    \item \textbf{Assisted, not autonomous reading}: the interpretation of technical documents (identification of indications, contraindications, restrictions, warnings, evidence, and normative silences) is performed by a technical reading environment that combines computational capabilities with explicit human curation. The computational model assists in document organization and decomposition, but never assumes normative or clinical authority.

    \item \textbf{Governed contextual presentation}: no assertion is presented to the end user outside a valid context of authority, scope, purpose, and risk. The system does not answer ``what is true'' but ``what is valid to assert here and now,'' distinguishing with precision regulated indication, evidence-supported off-label use, authorized orphan use, critical restriction, or simple absence of basis.

    \item \textbf{Integral traceability}: every assertion presented to the end user is accompanied by formal evidence trails, including identification of the primary source, document version, specific text nodes from which the information was extracted, and the interpretive trajectory that led to the assertion.

    \item \textbf{Human and institutional responsibility}: the system does not decide, does not prescribe, and does not authorize. It preserves sources, organizes the interpretive space, and documents the knowledge construction trajectory, keeping responsibility firmly in the human and institutional domain.
\end{enumerate}

Collectively, these five principles constitute what may be termed a \emph{governance-first architecture}: one in which provenance integrity, curatorial validation, and bidirectional traceability are primary design constraints rather than downstream compliance requirements \cite{Haarbrandt2018}. In contrast to output-oriented systems where governance is retrofitted as an institutional overlay, here governance is the organizing principle from which all downstream representations are derived.

The central \emph{breakthrough} of this architecture is not technological but epistemological: the model enables informing complex health decisions without usurping the regulatory role or clinical judgment, something few systems in the world can do coherently, defensibly, and at scale. This position finds direct grounding in the sociotechnical systems literature, which demonstrates that human action is always situated, contextual, and partially unpredictable, rendering the full delegation of authority to automated systems inadequate \cite{Suchman1987,Harrison2007}. Recent work confirms that this concern has intensified with the emergence of large language models, arguing that technology-centered approaches render AI systems fundamentally incompatible with clinical practice when they attempt to replicate rather than support the epistemic functions of health professionals \cite{Sokol2025}.

\subsection{PATOS: sovereign documentary memory}

PATOS (\emph{Preservation and Traceability of Official Sources})\footnote{The name also alludes to \emph{Lagoa dos Patos}, the largest lagoon in Brazil, located in the state of Rio Grande do Sul, where this work originated. Its primary function is the ingestion, preservation, versioning, and provision of regulatory, clinical, and normative documents in their original integrity, without anticipating interpretations or semantic transformations.} constitutes the documentary preservation layer of the architecture, conceived as a trusted digital repository oriented by the principles of the OAIS model \cite{CCSDS2012} and the \emph{Trusted Digital Repositories} guidelines \cite{OCLCCRL2007}.

The current PATOS implementation is organized as a structured documentary \emph{data lake} in four progressive maturity stages: \textbf{RAW} (original document, preserved bit-for-bit), \textbf{CLEANED} (format and encoding normalization, without content alteration), \textbf{STRUCTURED} (metadata extraction, section identification, and indexing), and \textbf{CURATED} (supervised enrichment and linking to canonical entities). Each document receives a unique identifier, cryptographic \emph{checksums} (SHA-256), and provenance metadata that enable the complete reconstruction of the ingestion and transformation history.

The repository aggregates documents from multiple Brazilian regulatory and institutional sources, including the National Health Surveillance Agency (ANVISA) and reference hospital databases, encompassing professional and patient package inserts, monographs, summary of product characteristics (\emph{SmPCs}), clinical protocols, and normative acts. The first operational version of the system contains more than 17,000 pharmaceutical documents, covering the entirety of reference medications registered in Brazil, in compliance with the documentary requirements of RDC (Resolu\c{c}\~{a}o da Diretoria Colegiada) 47/2009.

The central architectural principle of PATOS is RAW layer immutability: once ingested, the source document is never altered or replaced. Successive versions of the same document coexist in the repository, enabling historical reconstruction and temporal analysis of regulatory changes. This design decision, which distinguishes PATOS from databases that overwrite previous records, ensures that every subsequent interpretation can be verified against the exact source upon which it was built.

\subsection{Lector: assisted technical reading and evidential memory}

Lector represents the second functional stage of the architecture, positioned as an environment for machine-assisted technical reading and explicit human curation. Its conception draws on the tradition of \emph{computational sensemaking} \cite{PirolliRussell2011} and \emph{exploratory search} \cite{Hearst2009}, treating the interpretation of regulatory documents not as an automatic extraction operation but as an iterative cognitive process that combines computational decomposition with qualified professional judgment.

Lector's operational chain articulates four formal concepts:

\begin{enumerate}
    \item \textbf{DocumentRef}: reference to the document preserved in PATOS, including identifier, version, provenance metadata, and integrity \emph{checksum}. This reference ensures that every reading operation is anchored in a verifiable and immutable source.

    \item \textbf{PageIndex}: hierarchical decomposition of the document into navigable units, produced by language models (the architecture is model-agnostic; the current implementation uses Mistral 7B) under parameterized supervision. The \emph{PageIndex} generates a tree of sections and subsections with semantic summaries, enabling non-linear navigation and precise location of relevant content without the need for full document reading.

    \item \textbf{Evidence Pack}: formal evidence packaging unit, described in detail in subsection~3.6. Each \emph{Evidence Pack} encapsulates a qualified question, a grounded response, references to primary sources (including identifiers of specific \emph{PageIndex} nodes), explicit epistemic limits, and curation metadata.

    \item \textbf{CuratorialDecision}: formal record of the human decision on an \emph{Evidence Pack}, including acceptance, rejection, or revision request, with documented justification. This artifact materializes the principle that interpretive authority remains in the human domain, and that the computational model functions as a reading assistant, not as a normative authority.
\end{enumerate}

Lector operates according to an explicit internal norm: \emph{the model learns exclusively from already-legitimated human curation}. This means that the computational system never produces autonomous interpretations that become reference; every assertion that enters the system's body of knowledge has passed through documented human validation. This deliberate restriction, which may seem limiting from a scalability standpoint, is in reality the mechanism that guarantees the system's institutional reliability in a domain where interpretive errors can have severe clinical, legal, and regulatory consequences.

Lector identifies and qualifies distinct types of assertions present in regulatory documents, including: regulated indication, absolute or relative contraindication, dosing, drug interaction, warning, precaution, use in special populations, and normative silence (the explicit absence of regulatory pronouncement on a clinically relevant question). This typification is fundamental because the strength and scope of a pharmaceutical assertion depend critically on its normative nature (each type corresponds to a distinct illocutionary act, in Searle's terminology \cite{Searle1979}), and collapsing these distinctions into a generic ``information'' category is precisely the kind of simplification that generates risks in decision support systems. In particular, the explicit treatment of normative silence as a formal category, and not as mere absence of data, is consistent with the concept of \emph{normative gap} formalized in the deontic logic of normative systems \cite{AlchourronBulygin1971,Navarro2014,Prakken2015}. In this framework, the indeterminacy of a normative system does not equate to permission or prohibition but constitutes an epistemically relevant condition that should be communicated as such.

\subsection{The RPDA framework: contextual refraction of pharmaceutical information}

RPDA (Regulatory, Prescription, Dispensing, and Administration) constitutes the central information organization framework in PRISMA's presentation layer. Its contribution lies not in describing a care process (a role already fulfilled by the \emph{Medication-Use Process} consolidated in the literature \cite{StultzNahata2012}), but in formalizing a \emph{contextual presentation} principle: the same pharmaceutical information should be organized, filtered, and presented differently according to the decision-making context, the professional profile, and the risk associated with the action.

This distinction between \emph{process model} and \emph{information architecture} is central to understanding the proposal. The MUP describes \emph{what happens} with the medication throughout the care cycle; RPDA describes \emph{how information about the medication should be structured and presented} to support professional decisions at each phase of that cycle, without collapsing semantics, governance, or traceability.

The framework operates on three formal concepts:

\textbf{Informational core}: a structured, canonical, and versionable set of data that describes the medication independently of any specific use case, maintaining traceability and semantic consistency over time. In the context of the proposed architecture, the informational core is constituted by the documents preserved in PATOS and the qualified semantic units produced by Lector.

\textbf{Informational view}: a contextual projection of the informational core, organized according to a specific professional purpose, in which only the attributes, evidence, and assertions relevant to that decision are exposed, with granularity and governance compatible with the risk involved.

\textbf{Risk-oriented granularity}: the principle that the level of informational detail presented should be proportional to the potential impact of the supported decision, avoiding both critical omission and irrelevant information excess, a phenomenon widely documented in the literature as a source of \emph{alert fatigue} and clinical attention degradation \cite{StultzNahata2012,Brown2016}.

Each RPDA axis defines a distinct decision-making context:

\begin{itemize}
    \item \textbf{Regulatory}: information is organized to meet formal requirements of registration, compliance, traceability, and surveillance. It prioritizes official identifiers, version history, regulatory status, normative links, and legal framing. The prototypical professional is the regulatory pharmacist or the regulatory affairs manager.

    \item \textbf{Prescription}: information is reorganized to support the therapeutic decision, emphasizing indications, pharmaceutical forms, concentrations, routes of administration, use restrictions, and relevant clinical evidence, with semantic clarity and ambiguity reduction. The prototypical professional is the prescribing physician or the clinical pharmacist in reconciliation activity.

    \item \textbf{Dispensing}: information assumes an operational character, highlighting marketable presentations, packaging, fractionation, therapeutic equivalences, and safe substitution possibilities, minimizing logistic risks and supply errors. The prototypical professional is the dispensing pharmacist or the hospital pharmacy technician.

    \item \textbf{Administration}: the informational view is refined to support safe medication preparation and treatment execution, encompassing dosing, reconstitution, dilution, stability, storage, route-specific preparation instructions, and practical administration guidance. The prototypical professionals are the pharmacist responsible for medication preparation and the nurse responsible for administration.
\end{itemize}

RPDA does not create new data but organizes and exposes existing information situationally, aligning content, context, and professional responsibility. This need for differentiated treatment finds grounding in the theory of communicative action, as applied to information systems by Ngwenyama and Lee \cite{Ngwenyama1997} and by Mingers \cite{Mingers1995}, according to which every communicative act oriented toward understanding simultaneously raises claims of truth, normative rightness, and sincerity \cite{Habermas1984}: each RPDA axis prioritizes distinct validity claims, appropriate to the decision-making context and the risk involved. Additionally, speech act theory \cite{Austin1975,Searle1979} grounds the need to typify pharmaceutical assertions by their illocutionary force (assertives, directives, commissives), a typification that RPDA operationalizes by organizing each view according to the predominant nature of assertions relevant to that context.

Functionally, RPDA operates as the \emph{refraction} mechanism through which PRISMA projects the same informational core into distinct views, in deliberate analogy with the optical phenomenon in which a beam of light, upon traversing a prism, is decomposed into its components without the original light being altered or destroyed.

\subsection{PRISMA: contextual presentation and context graphs}

PRISMA (\emph{Pharmaceutical Refraction and Information System for Medication Assurance}) constitutes the contextual presentation layer of the architecture, responsible for organizing pharmaceutical knowledge into versioned assertions accompanied by formal evidence trails, and projecting it into the informational views defined by the RPDA framework.

\begin{figure}[ht]
\centering
\includegraphics[width=\textwidth]{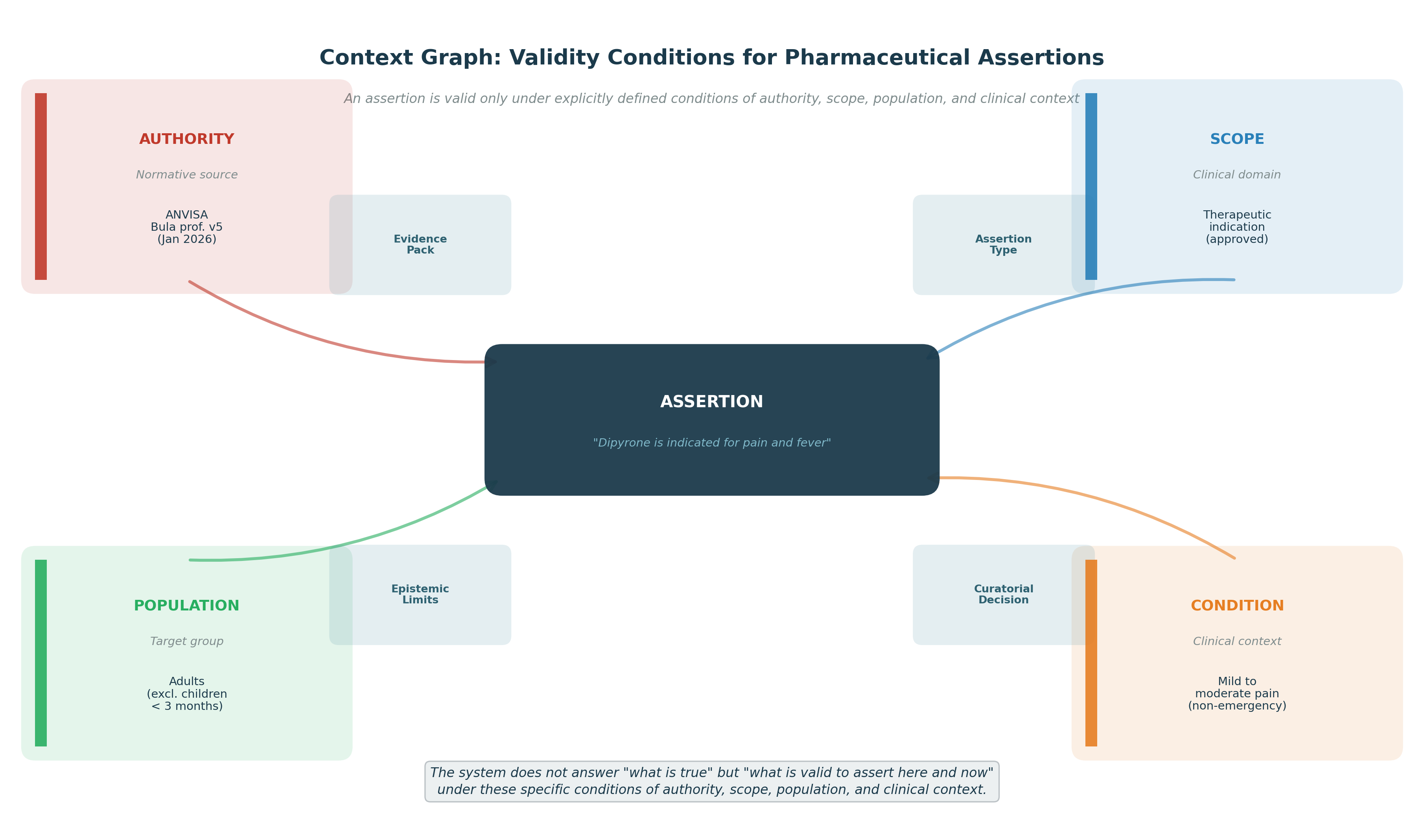}
\caption{Context graph structure illustrating the four validity dimensions for pharmaceutical assertions. Each assertion is valid only under explicitly defined conditions of authority (normative source), scope (clinical domain), population (target group), and clinical context. The example shows the assertion ``Dipyrone is indicated for pain and fever'' with its specific validity conditions. Qualifier nodes (Evidence Pack, Assertion Type, Epistemic Limits, Curatorial Decision) mediate between the assertion and its contextual dimensions.}
\label{fig:context-graphs}
\end{figure}

PRISMA's design is grounded in \textbf{context graphs}, structures that explicitly represent the conditions under which a pharmaceutical assertion is valid, including the normative source that supports it, the applicable regulatory scope, the target population, relevant clinical conditions, qualifiers that modulate its strength (e.g., ``in adults,'' ``except in severe hepatic insufficiency,'' ``per institutional protocol''), and recognized epistemic limits (Figure~\ref{fig:context-graphs}). This approach echoes Br\'{e}zillon's \emph{contextual graphs} \cite{Brezillon2005} and Toulmin's computational argumentation models \cite{Toulmin1958}, in which the validity of an assertion is not absolute but conditional on explicit \emph{warrants}, \emph{qualifiers}, and \emph{rebuttals}.

Concretely, when a health professional queries PRISMA about a medication, the system does not return a generic, undifferentiated answer, but a contextualized presentation that:

\begin{itemize}
    \item Identifies the active RPDA axis (Regulatory, Prescription, Dispensing, or Administration) and adjusts the corresponding informational view;
    \item Selects the relevant \emph{Evidence Packs} for that context, filtering assertions by type (indication, contraindication, dosing, warning, etc.) and by pertinence to the decision scenario;
    \item Presents each assertion accompanied by its provenance chain: source document, version, specific section, assertion type, and curatorial status;
    \item Explicitly signals epistemic limits: divergences between sources, documentary gaps, clinical context dependencies, and areas of normative silence.
\end{itemize}

The current PRISMA prototype, in implementation phase, is organized into five interface modules corresponding to the RPDA axes: \textbf{Search} (cross-cutting search), \textbf{Regulatory} (compliance and traceability), \textbf{Prescribe} (therapeutic decision support), \textbf{Dispense} (operations and equivalences), and \textbf{Administer} (safe execution). Each module presents the same informational base under the appropriate view, demonstrating in practice the principle of contextual refraction.

\subsection{Evidence Pack: formal unit of accountable assertion}

The \emph{Evidence Pack} concept constitutes the central articulation mechanism between the three architectural layers. An \emph{Evidence Pack} is defined as a formal pharmaceutical evidence packaging unit, composed of:

\begin{itemize}
    \item \textbf{Qualified question}: a typified formulation of the question the \emph{pack} answers, classified according to a taxonomy of pharmaceutical question types (indication, contraindication, dosing, interaction, use in special populations, among others). Question typification is essential because it determines the expected response type, applicable validation criteria, and associated governance.

    \item \textbf{Grounded response}: an assertion constructed from the technical reading of the source document, written in professional language, accompanied by explicit qualifiers that delimit its scope and validity conditions.

    \item \textbf{Provenance chain}: traceable references to the source document preserved in PATOS, including document identifier, version, \emph{checksum}, and identifiers of the specific \emph{PageIndex} nodes from which the information was extracted, enabling any assertion to be verified against the exact source that supports it.

    \item \textbf{Epistemic limits}: an explicit declaration of recognized limitations, including: divergences between distinct sources on the same topic, documentary gaps (information not available or not located), clinical context dependencies that affect assertion validity, and areas of normative silence, that is, situations in which the regulatory source does not explicitly pronounce on a clinically relevant question.

    \item \textbf{Curatorial status}: a record of the \emph{pack}'s validation stage in the curation workflow (draft, under review, accepted, rejected), with identification of the responsible curator and documented justification.
\end{itemize}

The \emph{Evidence Pack} materializes, at the operational level, the separation between data, interpretation, and presentation that grounds the entire architecture, operationalizing in the pharmaceutical domain the evidence-based medicine requirement that every clinical decision must be supported by explicit evidence, critically evaluated and applied in a contextualized manner \cite{Sackett1996}. The formal distinction between evidence quality and recommendation strength, established by the GRADE framework \cite{Guyatt2008}, is incorporated into the \emph{Evidence Pack} through the separation between the grounded response (what the evidence supports) and the epistemic limits (with what confidence and under what conditions).

The \emph{Evidence Pack} is not a ``system response''; it is an auditable unit of pharmaceutical knowledge, whose construction is traceable, whose validation is documented, and whose limits are explicit. This property fundamentally distinguishes it from responses generated by RAG systems or conversational agents, in which information provenance is typically irrecoverable after generation.

\subsection{Informational flow and architectural integration}

\begin{figure}[ht]
\centering
\includegraphics[width=\textwidth]{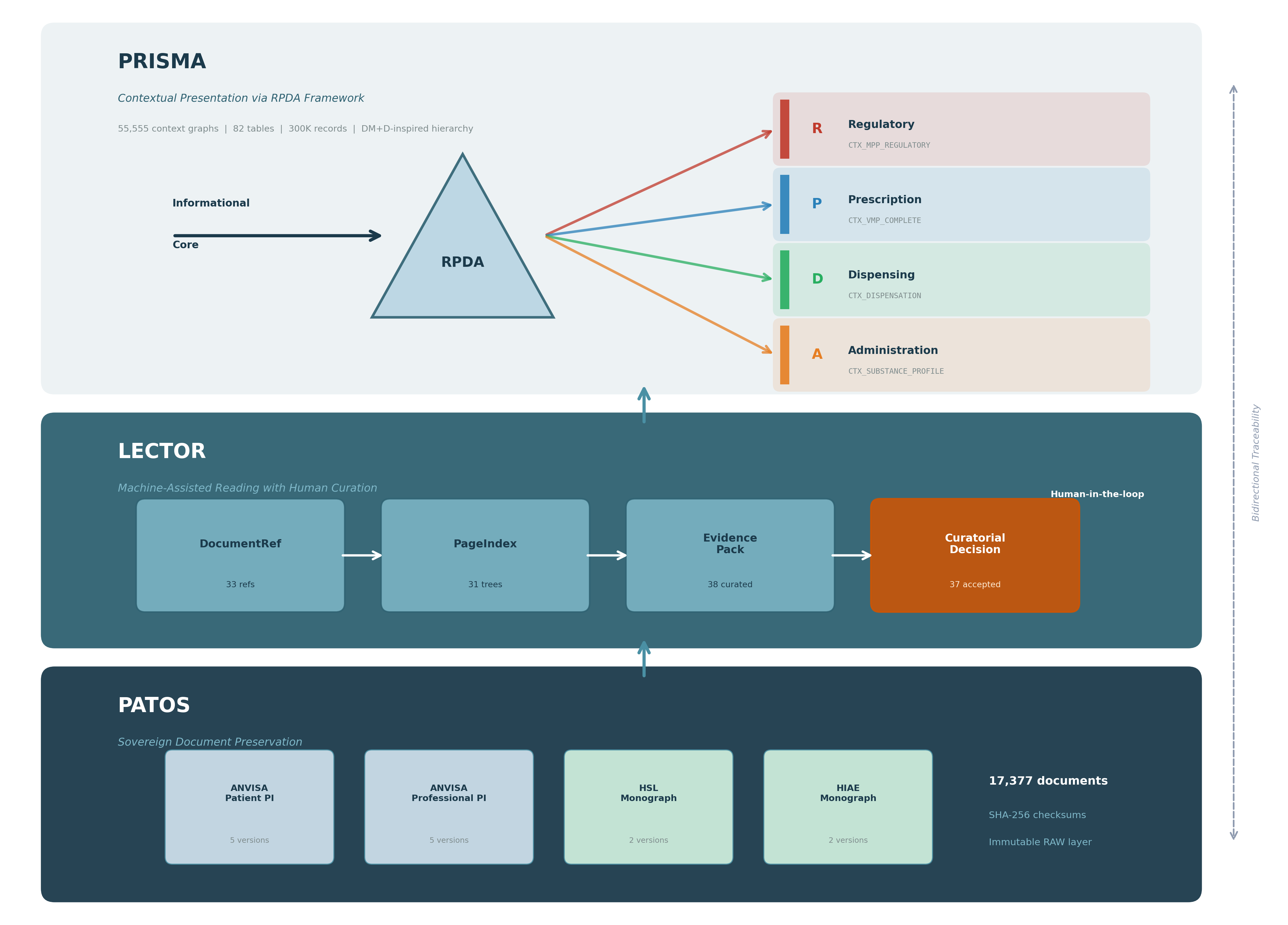}
\caption{Architectural overview of the PATOS--Lector--PRISMA infrastructure. The bottom layer (PATOS) preserves regulatory and institutional documents with immutability, versioning, and cryptographic checksums. The middle layer (Lector) implements machine-assisted reading, producing Evidence Packs validated by human curators. The top layer (PRISMA) refracts the informational core through the RPDA framework into four contextual views (Regulatory, Prescription, Dispensing, Administration). Solid arrows indicate unidirectional information flow; dashed arrows indicate bidirectional traceability from any assertion back to its source document.}
\label{fig:plp-architecture}
\end{figure}

Figure~\ref{fig:plp-architecture} provides an architectural overview of the three-layer infrastructure and its informational flow. The integration between the three layers follows a unidirectional and auditable informational flow:

\begin{enumerate}
    \item \textbf{Ingestion and preservation} (PATOS): regulatory and clinical documents are ingested, preserved in their original integrity, versioned, and indexed. The RAW layer remains immutable; subsequent transformations are recorded with complete provenance.

    \item \textbf{Reading and curation} (Lector): preserved documents are decomposed into navigable trees (\emph{PageIndex}), technically read by a combination of computational processing and human curation, and transformed into qualified \emph{Evidence Packs}. The computational model assists reading; the human validates interpretation.

    \item \textbf{Refraction and presentation} (PRISMA/RPDA): curated \emph{Evidence Packs} are organized into context graphs and projected into the informational views defined by RPDA, according to the active axis, the professional profile, and the associated risk. Presentation is always accompanied by provenance trails and explicit epistemic limits.
\end{enumerate}

This flow ensures that no information presented to the end user exists without anchoring in a verifiable source, without passage through a documented reading stage, and without explicit contextualization. The inversion or short-circuiting of this flow (for example, generating responses directly from documents without a curation stage, or presenting assertions without traceable provenance) is treated as an architectural violation, not as an acceptable simplification.

\noindent From the standpoint of implementation maturity, it is important to register with transparency that the three layers are at different degrees of maturity: PATOS is operational, with an active repository containing more than 17,000 documents; Lector is in active development, with its first two phases complete and the third in progress; and PRISMA is in high-fidelity prototyping phase, with developed user interfaces that demonstrate the principles of contextual refraction, but without complete backend integration. This transparency is deliberate: the paper proposes a conceptual architecture and its guiding principles, grounded in established academic traditions and partially validated by implementation, without claiming a complete and finalized system.

\subsection{Worked example: dipyrone monohydrate across the three layers}

To make the flow described in the preceding sections tangible, this subsection traces a concrete medication, dipyrone monohydrate (Novalgina, ANVISA registration no.\ 186200018), through the three architectural layers, using real data from the operating system.

The choice of dipyrone (metamizole) as an illustrative case is deliberate and epistemologically significant. It is a medication widely used in Brazil as a first-line analgesic and antipyretic, available without prescription, with more than 215 million doses administered annually and consumption of 488 tonnes in the state of S\~{a}o Paulo alone in 2016 \cite{Melo2023}. Paradoxically, dipyrone is banned in the United States, the United Kingdom, and France, and restricted in other European countries, due to its association with agranulocytosis. However, large-scale pharmacovigilance data from Brazil, including a retrospective study with 384,668 patients using real-world data, identified no cases of agranulocytosis in the cohort and reported that adverse effects of dipyrone were 38.8\% less frequent than those of paracetamol and 46.8\% less frequent than those of acetylsalicylic acid \cite{Sznejder2022}. Multicenter Latin American studies estimate the incidence of metamizole-associated agranulocytosis at 0.38 cases per million inhabitants-year, significantly lower than the historical estimates that motivated the ban in other countries \cite{LATIN2005}.

This regulatory divergence between jurisdictions makes dipyrone a paradigmatic case for the proposed architecture: pharmaceutical information about this medication is radically dependent on the normative context in which it is produced and interpreted. An infrastructure that uncritically imports international knowledge bases would classify dipyrone as a high-risk or prohibited substance, invalidating decades of Brazilian pharmacoepidemiological evidence and ANVISA's sovereign regulatory decision. PATOS, by preserving Brazilian regulatory documents with explicit provenance, and Lector, by producing Evidence Packs anchored in these national sources, ensure that the pharmaceutical knowledge presented by PRISMA reflects the regulatory and epidemiological context in which clinical decisions are actually made.

\subsubsection{Layer 1, PATOS: document preservation}

The PATOS repository contains 192 documents related to dipyrone, from four independent sources: official ANVISA package inserts (12 documents, including patient and professional versions), hospital monographs from Hospital S\'{i}rio-Liban\^{e}s and Hospital Albert Einstein (12 documents), and public pharmaceutical information databases (168 documents covering 170 commercial presentations from multiple manufacturers).

For the Novalgina registration (no.\ 186200018), PATOS preserves five complete historical versions of the package insert, spanning from February 2024 to January 2026, each with patient/professional pairs. The current version (\texttt{186200018\_20260116\_pac.pdf}, SHA-256 checksum: \texttt{d55ebcaf\ldots a7f6}) coexists with previous versions, without overwriting them. This preservation enables, for example, tracking when a contraindication was added or when the wording of a warning was modified by ANVISA, reconstructing the medication's regulatory evolution over time.

Each document receives a unique identifier (\texttt{DocumentRef}) at the moment of ingestion. For the current Novalgina professional package insert, the \texttt{DocumentRef} records: source (ANVISA), type (professional package insert), medication (novalgina), active ingredient (dipyrone monohydrate), format (PDF), capture date (2026-01-28), SHA-256 checksum, and currency status (\texttt{is\_current = 1}). Previous versions are marked as \texttt{is\_current = 0} but remain fully accessible in the repository.

\subsubsection{Layer 2, Lector: assisted reading and Evidence Packs}

From the document preserved in PATOS, Lector executes the technical reading chain in three stages.

\textbf{PageIndex.} The language model (currently Mistral 7B via Ollama, though the pipeline is model-agnostic) decomposes the professional package insert into a hierarchical tree of sections. For the Novalgina package insert, this decomposition produces nodes such as:

\begin{itemize}
    \item Node 1.1: ``What is this medication indicated for?'' (therapeutic indications)
    \item Node 1.3: ``When should I not use this medication?'' (contraindications)
    \item Node 1.5: ``How should I use this medication?'' (dosing and administration)
    \item Node 1.7: ``What adverse effects may this medication cause?'' (adverse reactions)
\end{itemize}

\noindent Each node receives a model-generated semantic summary, enabling non-linear navigation and precise location of relevant content.

\textbf{Evidence Pack.} For each clinically relevant question type, Lector generates a structured \emph{Evidence Pack}. Consider the Evidence Pack of type \texttt{INDICATION} for Novalgina:

\begin{lstlisting}
Evidence Pack EP-001 (INDICATION)
+-- Question: "What is this medication indicated for?"
+-- Focus: dipyrone monohydrate 500 mg tablet
+-- Findings:
|   +-- Approved indication: "Analgesic" (adults)
|   +-- Approved indication: "Antipyretic" (adults)
+-- Validity conditions: "Mild to moderate pain"
+-- Invalidity conditions: "Hypersensitivity to dipyrone"
+-- Traceability:
|   +-- Source: ANVISA, professional package insert
|   +-- Document: DocumentRef -> 186200018_20260116_prof.pdf
|   +-- PageIndex node: 1.1
|   +-- Document date: 2026-01-16
+-- Epistemic limits:
|   +-- Not covered: "Off-label indications"
|   +-- Context-dependent: "Always consult a healthcare professional"
+-- Curatorial status: ACCEPTED (curator: [identifier], date: [timestamp])
\end{lstlisting}

\noindent This same process is repeated for each assertion type (contraindication, dosing, interaction, adverse reaction), generating a set of Evidence Packs that together constitute the medication's curated evidential memory. In the current implementation, the system contains 38 curated Evidence Packs for the five reference medications, with 37 accepted and 1 rejected, linked to 119 canonical links to the PRISMA ontology.

\textbf{Normative silence as a formal category.} Not all clinically relevant questions receive an explicit answer from regulatory sources. The Novalgina professional package insert, for example, does not pronounce on dipyrone use in patients with glucose-6-phosphate dehydrogenase (G6PD) deficiency, a condition known to be clinically relevant for other analgesics and antipyretics. In the PLP architecture, this absence is not ignored but formally captured as an Evidence Pack of type \texttt{NORMATIVE\_SILENCE}:

\begin{lstlisting}
Evidence Pack EP-039 (NORMATIVE_SILENCE)
+-- Question: "Is dipyrone safe for patients
|   with G6PD deficiency?"
+-- Focus: dipyrone monohydrate 500 mg tablet
+-- Finding: No regulatory pronouncement identified
+-- Traceability:
|   +-- Source: ANVISA, professional package insert
|   +-- Document: DocumentRef -> 186200018_20260116_prof.pdf
|   +-- Sections reviewed: 1.3 (contraindications),
|       1.4 (warnings)
+-- Epistemic limits:
|   +-- Silence does NOT equate to safety or permission
|   +-- Clinical judgment required; consult specialized literature
+-- Curatorial status: ACCEPTED (curator: [identifier], date: [timestamp])
\end{lstlisting}

\noindent This formalization of silence is architecturally significant: it transforms the absence of information from an invisible gap into an explicit, auditable, and actionable category, precisely the kind of nuance that probabilistic models tend to collapse (Section~2.7).

\textbf{Curatorial decision.} Each Evidence Pack undergoes documented human validation. The curator may accept, reject, or request revision, recording the justification. The principle is explicit: no assertion enters the system's body of knowledge without passing through human validation. This restriction ensures that provenance is both documentary (the source exists) and interpretive (the reading of the source was validated by a qualified professional).

\subsubsection{Layer 3, PRISMA: canonical ontology and RPDA refraction}

Curated Evidence Packs are linked to PRISMA's pharmaceutical canonical ontology through \texttt{CanonicalLinks}, which materialize the bidirectional traceability between evidence and canonical entity. The ontology implements its own hierarchy, inspired by the DM+D (\emph{Dictionary of Medicines and Devices}, NHS, National Health Service, UK) model and expanded to accommodate the Brazilian regulatory reality, adopting universal identifiers and concepts from IDMP (Identification of Medicinal Products, ISO 11615/11616) for international interoperability. This adaptation is not a mere translation of foreign models, but a deliberate decision of sovereign governance: each level of the hierarchy receives a functional designation that reflects Brazil's regulatory, clinical, and logistic practice, while maintaining formal mappability with international standards. The canonical hierarchy, with 300,134 records distributed across 82 tables, is organized in five progressive levels of concretization:

\begin{lstlisting}
SUBSTANCE: dipyrone monohydrate (SUB-000033943)
  +-- CAS: 5907-38-0  |  UNII: 6429L0L52Y  |  DCB: 9564
  +-- 17 international identifiers (ChEMBL, DrugBank, PubChem, MeSH, ...)
  +-- 24 multilingual synonyms (dipyrone, metamizol, analgin, ...)
       v
VTM -- Qualitative Composition: dipyrone monohydrate (VTM-000010750)
  +-- 16 compositions contain this substance (mono + combinations)
       v
VMP -- Formulation: dipyrone monohydrate 500 mg tablet (VMP-000051605)
  +-- ATC: N02BB02
  +-- Pharmaceutical form: tablet (PDF-000002766)
  |   +-- EDQM taxonomy: oral / solid / ingestion / conventional release
  +-- Quantitative composition: 500 mg dipyrone monohydrate
  +-- 7 formulations available for this substance
       v
VMPP -- Pack: dipyrone 500 mg x 30 tablets (VMPP-000103766)
  +-- Prescribable unit: tablet
  +-- Primary packaging: blister
  +-- 13 virtual packs (x10, x20, x30, x50, x100, ...)
       v
AMP -- Product: NOVALGINA 500 mg (AMP, equiv. IDMP Medicinal Product)
  +-- Manufacturer: Sanofi Medley (ORG-000000032)
  +-- ANVISA registration: PMA 183260351
  +-- Therapeutic class: non-narcotic analgesics
       v
AMPP -- Trade Presentation: NOVALGINA 500 mg box x 30 (equiv. IDMP Packaged MP)
  +-- EAN: 7891058008635
  +-- Label: OTC (no prescription required)
  +-- 92 trade presentations from 24 manufacturers for this formulation
\end{lstlisting}

\noindent Table~\ref{tab:hierarchy} synthesizes the correspondence between canonical hierarchy levels, the DM+D model, and IDMP concepts:

\begin{table}[ht]
\centering
\caption{Correspondence between PRISMA canonical hierarchy levels, the DM+D model (NHS), and IDMP concepts (ISO 11615/11616).}
\label{tab:hierarchy}
\small
\begin{tabular}{@{}llll@{}}
\toprule
\textbf{PRISMA Level} & \textbf{Functional Designation} & \textbf{DM+D} & \textbf{IDMP (ISO 11615/11616)} \\
\midrule
SUBSTANCE & Substance & Substance & Substance \\
VTM & Qualitative Composition & Virtual Therapeutic Moiety & --- \\
VMP & Formulation & Virtual Medicinal Product & Pharmaceutical Product \\
VMPP & Pack & Virtual Medicinal Product Pack & --- \\
AMP & Product & Actual Medicinal Product & Medicinal Product \\
AMPP & Trade Presentation & Actual Medicinal Product Pack & Packaged Medicinal Product \\
\bottomrule
\end{tabular}
\end{table}

\textbf{RPDA refraction.} PRISMA projects this same informational base into distinct contextual views through five materialized \emph{context graphs} (55,555 JSON graphs in total, generated in under 0.45 seconds):

\begin{itemize}
    \item \textbf{Regulatory view} (\texttt{CTX\_MPP\_REGULATORY}): presents the trade product with ANVISA registration (PMA 183260351), authorization status, marketing date, label classification, therapeutic classification, regulatory category, storage conditions (15--30\textdegree C, 24-month shelf life), and EAN (European Article Number) for logistic traceability. This view serves the regulatory pharmacist who needs to verify compliance.

    \item \textbf{Prescription view} (\texttt{CTX\_VMP\_COMPLETE}): presents the virtual product with substance, concentration (500 mg), ATC (Anatomical Therapeutic Chemical classification, N02BB02), DDD (Defined Daily Dose, 0.167), pharmaceutical form with complete taxonomy (oral, solid, ingestion, conventional release), quantitative composition, and available presentations. This view serves the prescriber who needs to decide among pharmaceutical forms and concentrations.

    \item \textbf{Dispensing view} (\texttt{CTX\_DISPENSATION}): presents the virtual pack with prescribable unit (tablet), packaging detail (blister $\times$ 30), and lists the 17 trade products available in this configuration, with brand, manufacturer, EAN, label, and marketing status. This view serves the dispenser who needs to select among equivalent alternatives.

    \item \textbf{Administration view} (\texttt{CTX\_SUBSTANCE\_PROFILE}): presents the complete substance profile (17 international identifiers, 24 multilingual synonyms, DCB (Denomina\c{c}\~{a}o Comum Brasileira) code, and mapping to 16 VTMs and 36 VMPs), oriented toward safe medication preparation, administration, and therapeutic reconciliation. This view serves the pharmacist responsible for medication preparation, the nurse who needs to verify substance identity before administration, and the clinical pharmacist in therapeutic reconciliation.
\end{itemize}

\noindent Each of these views is derived from the same ontological base, which in turn is sustained by Lector's curated Evidence Packs, which in turn are anchored in the documents preserved in PATOS. The traceability chain is integral: from the contextual view presented to the professional to the specific node of the original regulatory document, passing through the Evidence Pack that formalized the interpretation and the curator who validated it.

\bigskip
\noindent\textbf{Summary of demonstrated capabilities.} Table~\ref{tab:metrics} summarizes the quantitative scope of the worked example, aggregating metrics across the three architectural layers for the dipyrone case.

\begin{table}[ht]
\centering
\caption{Quantitative summary of the worked example: dipyrone monohydrate across the three PLP layers.}
\label{tab:metrics}
\begin{tabular}{llr}
\toprule
\textbf{Metric} & \textbf{Layer} & \textbf{Value} \\
\midrule
Documents ingested & PATOS & 192 (4 sources) \\
Document types & PATOS & Inserts, monographs, databases \\
PageIndex trees generated & Lector & 31 \\
Evidence Packs curated & Lector & 38 (37 accepted, 1 rejected) \\
Canonical links & Lector $\rightarrow$ PRISMA & 119 \\
Ontology records & PRISMA & 300,134 (82 tables) \\
Hierarchy levels & PRISMA & 6 (SUBSTANCE $\rightarrow$ AMPP) \\
Context graphs generated & PRISMA/RPDA & 55,555 (in $<$ 0.45\,s) \\
RPDA views demonstrated & PRISMA & 4 \\
Assertion types formalized & Lector & 7 \\
\bottomrule
\end{tabular}
\end{table}

\section{Motivating Case: Complementarity with Operational Decision Support Systems}

The preceding section presented the PATOS--Lector--PRISMA architecture as a conceptual proposal, grounded in established academic traditions and partially validated by implementation. To demonstrate the practical relevance of this proposal, this section examines how the infrastructure would complement, without replacing, operational pharmaceutical decision support systems already in production, taking as a motivating case NoHarm.ai, a Brazilian system widely adopted in hospital pharmacy \cite{DosSantos2019,Leitao2023}.

\subsection{What operational systems do well}

Systems such as NoHarm.ai operate in the \emph{real-time risk detection and mitigation} layer: they analyze prescriptions against knowledge bases, identify drug interactions, out-of-range doses, therapeutic duplicities, incompatibilities, and high-risk profiles, and generate alerts that guide pharmaceutical intervention. When well-calibrated, these systems substantially amplify the clinical pharmacist's professional capacity, enabling limited teams to cover a prescription volume that would be humanly impossible to review manually.

This is a legitimate and clinically demonstrated operational contribution. However, it operates on an implicit premise: \emph{that the underlying knowledge base is reliable, current, contextualized, and traceable}. It is precisely at this premise that the proposed architecture intervenes.

\subsection{Structural gaps that persist}

Even the best operational pharmaceutical decision support systems share a set of structural limitations that stem not from engineering failures, but from the absence of underlying informational layers:

\textbf{Absence of traceable documentary provenance.} When a system issues a drug interaction alert, the professional typically has no access to the regulatory source that supports that alert: the specific package insert, the document version, the exact section, the original context of the assertion. The alert exists as a computational fact unanchored from its primary evidence. If the professional needs to verify, justify, or question the alert, they must resort to external sources, manually reconstructing the provenance chain that the system should preserve.

\textbf{Opacity of the interpretive trajectory.} Between the original regulatory document and the alert presented to the professional, there exists an interpretation trajectory (information extraction, terminological normalization, severity classification, rule application) that remains typically invisible. The professional receives the final result (the alert) without access to the reasoning that produced it. In a domain where institutional trust depends on auditability, this opacity represents a systemic fragility, not an acceptable simplification.

\textbf{Contextual indifferentiation.} The same pharmaceutical information (for example, a relative contraindication) can have radically different relevance depending on the decision-making context: in the regulatory axis, it may require formal registration; in the prescriptive axis, it may demand risk-benefit assessment; in the dispensing axis, it may not be relevant; in the administration axis, it may generate a specific monitoring instruction. Systems that present information in an undifferentiated manner (the same alert, in the same format, for all contexts) miss the opportunity to tailor the presentation to the risk and responsibility involved.

\textbf{Binary treatment of knowledge.} Alert systems operate predominantly in binary logic: the interaction exists or does not exist; the dose is adequate or inadequate; the allergy is registered or not. This approach does not naturally accommodate the nuances that characterize regulatory pharmaceutical knowledge: evidence-supported off-label indications, context-dependent restrictions, divergences between distinct regulatory sources, or normative silences that do not equate to permissions. The 63\% to 95\% \emph{override} rate documented in the literature \cite{StultzNahata2012} may be, in part, a consequence of this inadequately binary treatment of knowledge that is inherently graduated and contextual.

\subsection{How the PATOS--Lector--PRISMA infrastructure would complement the operational cycle}

The proposal is not to replace operational systems such as NoHarm.ai, but to provide them with an underlying informational infrastructure that resolves the gaps identified above. The complementarity can be articulated along four dimensions:

\textbf{Documentary anchoring.} With PATOS as the preservation layer, every alert generated by an operational system could be linked to the specific regulatory source (document, version, section) that supports the rule or information used. The professional receiving an alert would gain direct access to the primary evidence, without manual reconstruction.

\textbf{Interpretive transparency.} With Lector as the technical reading layer, the trajectory between the source document and the applied rule would be documented in auditable \emph{Evidence Packs}: what question was posed to the document, what response was extracted, what epistemic limits were identified, and who validated the interpretation. The system's opacity would be replaced by a reconstructible interpretive trajectory.

\textbf{Contextual refraction.} With PRISMA and the RPDA framework, the same pharmaceutical information would be presented differently according to the active decision axis. A system that today issues the same alert to the prescriber, the dispenser, and the nurse would instead contextualize the information according to the professional profile, the associated risk, and the granularity appropriate to the moment of decision.

\textbf{Graduated qualification.} With Lector's assertion typification and PRISMA's context graphs, the binary logic of alerts would be enriched by a more faithful representation of pharmaceutical knowledge: regulated indications vs.\ evidence-supported off-label use, absolute vs.\ relative contraindications, strong evidence vs.\ fragile consensus, and presence vs.\ explicit absence of regulatory pronouncement.

To make this complementarity concrete, consider a scenario already grounded in the data presented in Section~3.8: a prescriber orders dipyrone 500 mg for a patient concurrently receiving warfarin. An operational CDSS would issue a binary drug interaction alert. With the PLP infrastructure in place, that same alert would arrive enriched: (i) the specific Evidence Pack documenting the dipyrone--warfarin interaction (EP type: INTERACTION), anchored to the professional package insert v5 (PATOS document \texttt{BULA\_PROF\_NOVALGINA\_v5}, SHA-256 verified); (ii) the curatorial decision recording that the interaction is classified as \emph{moderate} with a recommendation to monitor INR, not as absolute contraindication; (iii) the RPDA-refracted presentation appropriate to the context, where the prescriber sees the pharmacological mechanism and clinical management recommendation, the dispenser sees the dispensing precaution, the nurse sees the monitoring parameters. The provenance chain from alert to source document would be fully reconstructible. This scenario, while not yet operationally validated through integration with an actual CDSS, uses exclusively real data entities from the operating system described in Section~3.8.

\subsection{Note on the nature of complementarity}

The complementarity described here is not additive (it is not about ``more features'' coupled to an existing system) but \emph{infrastructural}. The distinction is relevant: additional features operate on the same architectural plane as the system they complement; an infrastructure operates on an underlying plane, providing conditions of possibility that the operational system can exploit without needing to implement them internally.

In this sense, the relationship between PRISMA and a system like NoHarm.ai is analogous to the relationship between an operating system and an application: the application needs the operating system to function reliably, but does not replace it, nor is it replaced by it. The absence of a normative information infrastructure does not prevent operational systems from functioning (NoHarm.ai is proof of this), but structurally limits what these systems can offer in terms of traceability, explainability, and institutional accountability.

\section{Discussion}

\subsection{Architectural implications}

The proposal presented in this paper shifts the focus of the debate on AI in pharmacy, currently predominantly centered on algorithmic performance, model accuracy, and comparative benchmarks, toward a more fundamental discussion about \emph{information architecture} itself. This change of plane is not trivial: it implies recognizing that many of the limitations attributed to AI models or system interfaces are, in reality, consequences of an absent or inadequate informational infrastructure. This perspective finds convergent validation in analyses of the broader enterprise AI ecosystem. The MIT NANDA report \cite{ChallapallyMITNANDA2025} demonstrates that the bet that sufficiently advanced models would compensate for fragile data, brittle systems, and poorly defined processes has proved systematically false: for high-complexity work, 90\% of professionals still prefer humans, not out of generic distrust of AI but because current systems do not retain feedback, do not adapt to context, and do not improve over time. Notably, the five criteria that executives prioritize when evaluating AI systems (trust, deep workflow understanding, minimal disruption, clear data boundaries, and capacity for continuous improvement) align almost directly with the five design principles of the PLP architecture: documentary sovereignty, assisted reading with curation, governed contextual presentation, integral traceability, and human and institutional responsibility. Ramesh \cite{Ramesh2026}, commenting on these findings, proposes five diagnostic self-awareness tests for organizations, including the \emph{Campsite Test} (``did systems become healthier after AI introduction?'') and the \emph{Incentives Test} (``who is recognized for foundational work?''), which, transposed to the pharmaceutical domain, reinforce the thesis that the challenge is infrastructural, not algorithmic.

The formal separation between preservation (PATOS), interpretation (Lector), and presentation (PRISMA) is more than a modular organization of features; it reflects an epistemological distinction between three cognitive and institutional operations that possess distinct requirements of governance, temporality, auditability, and responsibility. The fusion of these operations (current practice in most existing systems) is not a benign simplification but an identifiable source of systemic fragilities: loss of provenance, interpretive opacity, \emph{alert fatigue}, and erosion of the boundaries of responsibility between machine, professional, and institution.

Convergently, Talisman \cite{Talisman2026} observes that the industry has invested extensively in \emph{semantic layers} for metric governance but may have solved the wrong problem: AI systems need \emph{context engineering} (concepts, relationships, constraints, and inference capabilities), not standardized calculations. The author notes that \emph{healthcare} and \emph{life sciences} have already demonstrated that formal ontologies work at scale, and that competitive advantage belongs to organizations that treat knowledge architecture as a core competency. In subsequent work, the same author \cite{Talisman2026b} deepens the diagnosis by arguing that LLMs destroy provenance structurally and irreversibly, dissolving the documentary chains that civilization has built over centuries to distinguish knowledge from mere assertion. The convergence of these two analyses, formulated independently from the domain of data engineering and metadata, doubly validates the central proposition of this work: the challenge of pharmaceutical information is simultaneously one of \emph{knowledge architecture} (what to represent and how to reason) and of \emph{provenance infrastructure} (how to preserve the chains that confer epistemic legitimacy to assertions). (We note that these sources \cite{Talisman2026,Talisman2026b,Ramesh2026} are practitioner-oriented analyses rather than peer-reviewed publications; we cite them for the convergence of their independently formulated diagnoses with the architectural thesis of this work, not as primary evidence.)

The RPDA framework's contribution, in this context, is to demonstrate that the problem is not limited to the vertical separation between layers but includes a horizontal dimension: the same information needs to be organized, filtered, and presented differently according to the decision-making context. The distinction between RPDA and the \emph{Medication-Use Process} (MUP) consolidated in the literature is central: while the MUP describes the medication flow in the care process, RPDA describes the \emph{information architecture about the medication}, proposing that each phase of the use cycle constitutes a distinct informational context, with its own requirements of granularity, governance, and risk.

\subsection{Relationship with existing approaches}

\begin{figure}[ht]
\centering
\includegraphics[width=\textwidth]{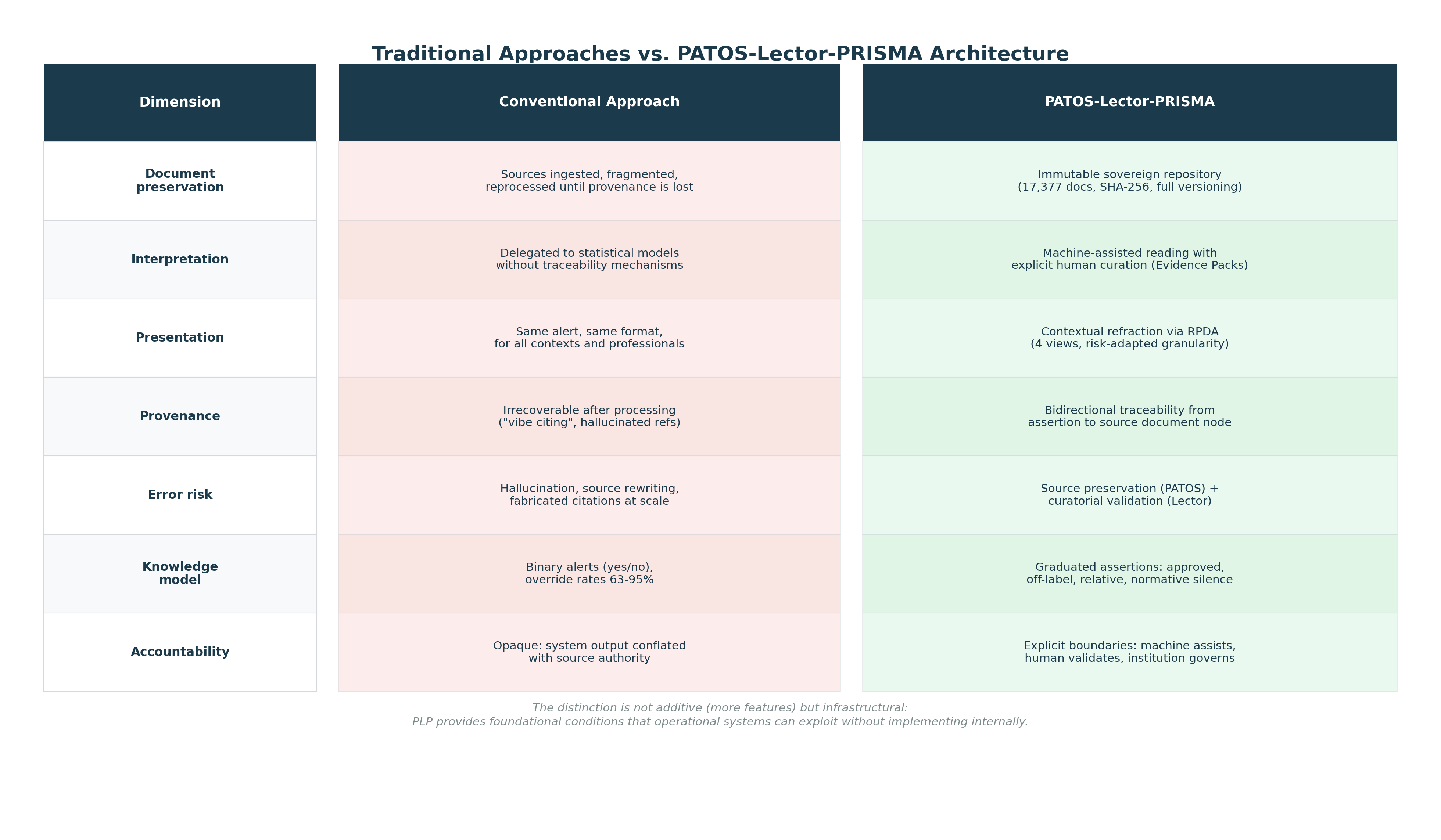}
\caption{Comparative analysis of conventional approaches versus the PATOS--Lector--PRISMA architecture across seven key dimensions: document preservation, interpretation, presentation, provenance, error risk, knowledge model, and accountability. The distinction is not additive (more features) but infrastructural: PLP provides foundational conditions that operational systems can exploit without implementing internally.}
\label{fig:comparison}
\end{figure}

It is important to situate the proposal in relation to contemporary approaches that share similar concerns, though operating on distinct planes. Figure~\ref{fig:comparison} provides a systematic comparison along seven key dimensions (document preservation, interpretation, presentation, provenance, error risk, knowledge model, and accountability), contrasting conventional approaches with the PLP architecture.

RAG systems, such as DrugRAG \cite{Kazemzadeh2025}, represent a significant advance by linking language model responses to documentary sources. However, ``retrieval'' in RAG is typically a technical vector search operation, not an assisted and curated technical reading act; and ``generation'' produces plausible text, not typified and traceable assertions. RAG partially solves the hallucination problem but not the interpretive accountability problem.

Pharmaceutical knowledge graphs, such as medicX-KG \cite{Farrugia2025}, offer structured representations of relationships between medications, indications, and interactions. These graphs, however, typically dissociate the represented knowledge from the documentary source that originated it, rendering provenance irrecoverable after graph construction. PRISMA's proposal is that context graphs should be \emph{derived} from preserved and curated sources, not constructed as autonomous entities.

Systems based on agents and normative YAMLs, such as that described by Sabel et al.\ \cite{Sabel2025}, demonstrate that even in highly controlled scenarios, the application of LLMs to regulated domains progressively requires the introduction of normative layers, flow control, human supervision, and formal specifications. This empirical convergence (authors who, without coordination, arrive at the same structural needs) functions as exogenous validation of the problem that the PATOS--Lector--PRISMA infrastructure proposes to solve systematically and reusably.

Bean et al.\ demonstrate in a controlled study with 1,298 participants that LLMs correctly identify medical conditions in 94.9\% of cases when tested in isolation, yet do not improve real users' performance: participants interacting with the same models scored below the control group ($p < 0.001$) \cite{Bean2026}. This dissociation between algorithmic competence and informational utility reinforces that the central problem is not one of model performance but of architecture of the mediation between knowledge and decision.

\subsection{Regulatory and institutional considerations}

The proposed architecture was designed with the Brazilian regulatory context as its horizon, particularly ANVISA's medication registration regime and the documentary requirements of RDC 47/2009 for package inserts. However, the underlying principles (documentary sovereignty, assisted technical reading, governed contextual presentation, and integral traceability) are conceptually jurisdiction-agnostic and applicable to any regulatory context involving normative documents, versions, professional interpretation, and decisions with clinical or legal consequences.

The deliberate non-decisional position of the system carries significant regulatory implications. In a scenario where global regulators are beginning to demand explainability and auditability from AI systems in health, an infrastructure that preserves sources, documents interpretive trajectories, and delineates responsibilities offers a more robust compliance model than systems operating as ``black boxes'' with plausible but untraceable outputs.

From an institutional standpoint, the \emph{Evidence Pack} concept introduces a unit of work and responsibility that is familiar to the pharmacist: a qualified question, a grounded response, identified sources, recognized limits, and a documented curatorial decision. This conceptual familiarity may facilitate institutional adoption of an infrastructure that, despite its architectural complexity, operates upon already-established professional practices.

\subsection{Evaluation criteria}

As a position paper proposing a normative architecture, this work does not claim empirical validation. However, to enable systematic assessment, both of the current proposal and of future implementations, we propose five evaluation criteria against which the PLP infrastructure should be judged:

\begin{enumerate}
    \item \textbf{Provenance completeness}: can every assertion presented to the end user be traced back to a specific source document, version, and text node?
    \item \textbf{Interpretive traceability}: is the trajectory from source to assertion (including the model's reading, the curator's decision, and the justification) fully reconstructible?
    \item \textbf{Curatorial coverage}: what proportion of clinically relevant assertions have passed through documented human validation?
    \item \textbf{Contextual differentiation}: does the system present the same information differently according to the professional context (RPDA axis), the associated risk, and the decision at hand?
    \item \textbf{Institutional accountability}: are the boundaries of responsibility between machine, professional, and institution explicitly delineated and auditable?
\end{enumerate}

\noindent The worked example in Section~3.8 provides preliminary evidence that the architecture satisfies criteria 1, 2, 4, and 5 for a single medication case. Criterion 3 is partially demonstrated (38 Evidence Packs curated for five reference medications). Systematic validation across a broader medication set, with controlled studies involving health professionals, constitutes the primary agenda for future empirical work.

\subsection{Limitations and future work}

The present work has limitations that must be explicitly acknowledged.

First, the proposal is fundamentally \emph{architectural and conceptual}. Although individual components are at different implementation stages (PATOS operational, Lector in development, PRISMA in prototyping), the complete integration between the three layers has not yet been demonstrated in a production environment. Empirical validation against the criteria defined in Section~5.4, including usability studies, clinical impact measurements, and scalability evaluations, constitutes a necessary and yet-unrealized research agenda.

Second, the explicit human curation the model requires, particularly in Lector's DocumentRef $\rightarrow$ PageIndex $\rightarrow$ EvidencePack $\rightarrow$ CuratorialDecision chain, implies operational costs and maturation times that may limit initial scalability. The question of how to scale curation without compromising reliability is an open challenge, though incremental strategies (prioritization by reference medications, learning from already-legitimated curation, progressive expansion by therapeutic families) offer viable paths.

Third, the RPDA framework proposes four axes of contextual refraction that, while grounded in professional practice and the \emph{Medication-Use Process} literature, may not exhaust all relevant decision-making contexts. Additional axes, such as pharmacovigilance, clinical research, or professional education, may require their own informational views, and the framework's extensibility to these contexts remains a matter for future investigation.

Fourth, the evaluation of language models as technical reading assistants (the role played by Lector's PageIndex) is rapidly evolving. The specific choice of models, \emph{prompting} strategies, and quality metrics for hierarchical decomposition of regulatory documents is a topic requiring dedicated investigation, and results may vary significantly as the underlying technology evolves.

Fifth, the architecture introduces operational risks that must be acknowledged. The curation overhead required by Lector's human-in-the-loop model may create latency incompatible with time-critical clinical scenarios, where immediate alerts take precedence over traceable interpretive trajectories. Additionally, despite the deliberate principle that the system does not decide, the structured and authoritative presentation of Evidence Packs may induce \emph{automation bias} (the tendency for professionals to accept system outputs uncritically), particularly under conditions of time pressure and cognitive load \cite{Ash2007}. Mitigating this risk requires explicit interface design strategies (such as uncertainty indicators, mandatory acknowledgment of epistemic limits, and friction mechanisms for high-stakes assertions) that remain to be tested empirically.

Sixth, although the current ontology already adopts IDMP concepts and identifiers (ISO 11615/11616) and a hierarchy inspired by the DM+D model (NHS), full interoperability with international standards for medication identification and clinical data exchange (such as FHIR, Fast Healthcare Interoperability Resources) is a natural direction of evolution, involving non-trivial challenges of terminological mapping, regulatory harmonization, and inter-institutional governance.

Finally, the empirical validation agenda must include medications of varying informational complexity: not only the five reference medications (dipyrone, ceftriaxone, clonazepam, warfarin, adalimumab), which were deliberately selected for their documentary richness and regulatory salience, but also common generic medications with fewer documentary sources, recently approved medications with limited post-marketing evidence, and medications with contested or evolving regulatory status. The architecture's robustness under conditions of sparse, conflicting, or rapidly changing information is a critical test that the current worked example, by design, does not address.

\section{Conclusions}

This paper presented the PATOS--Lector--PRISMA set as a proposal for a normative information infrastructure for responsible pharmaceutical knowledge management, grounded in the formal separation between documentary preservation, assisted interpretation, and governed contextual presentation.

The central contribution lies in the identification and treatment of a structural gap: most existing systems (whether RAG pipelines, conversational agents, alert systems, or rule engines) collapse into a single technical layer cognitive and institutional operations that possess radically distinct requirements of governance, temporality, and responsibility. This fusion, it is argued, constitutes the root cause of recurring fragilities such as loss of provenance, interpretive opacity, and erosion of the boundaries of responsibility between machine, professional, and institution.

The proposal articulates four specific contributions:

\begin{enumerate}
    \item \textbf{The PATOS--Lector--PRISMA triad} as an epistemologically grounded three-layer architecture, anchored in Buckland's tripartite distinction, the hermeneutic arc, and Floridi's levels of abstraction method.

    \item \textbf{The RPDA framework} (Regulatory, Prescription, Dispensing, and Administration) as a contextual refraction mechanism for pharmaceutical information, distinct from the \emph{Medication-Use Process} by operating on the plane of information architecture rather than on the plane of the care process.

    \item \textbf{The \emph{Evidence Pack} concept} as a formal unit of accountable assertion (versioned, traceable, typified, and accompanied by explicit epistemic limits), operationalizing in the pharmaceutical domain the principles of evidence-based medicine and computational argumentation.

    \item \textbf{The typification of pharmaceutical assertions} by their normative nature and illocutionary force, including the formal treatment of normative silence as an epistemically relevant category, grounded in speech act theory and deontic logic.
\end{enumerate}

The analysis of complementarity with operational systems such as NoHarm.ai (Section~4) demonstrates that the proposal does not compete with existing solutions but provides infrastructural conditions (documentary anchoring, interpretive transparency, contextual refraction, and graduated qualification) that extend what these systems can offer in terms of traceability and institutional accountability.

The work explicitly acknowledges that this is a conceptual and architectural proposal, with partial implementation and empirical validation yet to be conducted. However, it is argued that the clear formulation of the problem, rigorous epistemological grounding, and demonstration of feasibility through components under development constitute a relevant contribution to the field of health informatics, particularly at a time when the pressure for AI adoption in regulated domains demands trust infrastructures that most existing approaches do not provide.

The fundamental direction of this work can be synthesized in a single statement: \emph{the central challenge of pharmaceutical information is not algorithmic but architectural}. The response to this challenge demands not merely better models, but information infrastructures that preserve sources, document interpretations, contextualize presentations, and keep responsibility firmly in the human and institutional domain.


\section*{Author Contributions}

E.R.Z.N.\ conceived the PATOS--Lector--PRISMA architecture, designed the three-layer separation, developed the RPDA framework, led the system implementation, and wrote the manuscript. A.V.C.\ contributed to the design of the Evidence Pack model, the curatorial workflow, and the PRISMA ontology structure. C.C.\ contributed to pharmacovigilance analysis and regulatory document curation. G.P.G.\ contributed to the canonical hierarchy design and data population pipeline. B.M.\ contributed to the PATOS repository architecture and document ingestion infrastructure. All authors reviewed and approved the final manuscript.

\section*{Conflict of Interest Statement}

All authors are collaborators at Pharmdata, a startup developing the PRISMA system described in this paper. The PATOS--Lector--PRISMA infrastructure is the subject of active development by Pharmdata. The authors declare that the research was conducted in the absence of any commercial or financial relationships that could be construed as a potential conflict of interest beyond the stated affiliation.

\section*{Funding}

This research received no external funding. Development of the PATOS--Lector--PRISMA infrastructure is funded internally by Pharmdata.

\section*{Data Availability}

The data reported in Section~3.8 (worked example) are derived from publicly available regulatory documents published by ANVISA (National Health Surveillance Agency, Brazil). The PATOS repository, Lector curation artifacts, and PRISMA ontology are proprietary to Pharmdata. Aggregate statistics and architectural descriptions are provided in the manuscript to enable assessment of the proposal.

\bibliography{references}

\appendix

\section{Evidence Pack: Semi-Formal Specification}
\label{app:evidence-pack}

This appendix provides a semi-formal specification of the Evidence Pack, the central unit of accountable assertion in the PATOS--Lector--PRISMA architecture. The specification is intentionally conceptual, establishing the structural properties and well-formedness conditions that any conforming implementation must satisfy. A machine-readable formalization (e.g., in JSON-LD or RDF) is deferred to future work.

\subsection{Definition}

An Evidence Pack is a 5-tuple:

\begin{equation}
\mathbf{EP} = \langle Q, R, P, L, C \rangle
\end{equation}

\noindent where:

\begin{itemize}
    \item $\mathbf{Q}$ (Qualified Question) is a pair $(q, \tau)$ such that $q$ is a natural-language question and $\tau \in T$ is an assertion type drawn from a controlled taxonomy $T = \{\textsc{indication}, \textsc{contraindication}, \textsc{dosing}, \textsc{interaction}, \textsc{adverse\_reaction}, \textsc{warning}, \textsc{precaution}, \textsc{special\_population}, \textsc{normative\_silence}, \ldots\}$.

    \item $\mathbf{R}$ (Grounded Response) is a triple $(a, V^{+}, V^{-})$ such that $a$ is a natural-language assertion constructed from the technical reading of one or more source documents, $V^{+}$ is a set of validity conditions under which $a$ holds, and $V^{-}$ is a set of invalidity conditions under which $a$ does not hold.

    \item $\mathbf{P}$ (Provenance Chain) is a non-empty set of provenance records $\{p_1, p_2, \ldots, p_n\}$, where each $p_i = (d, v, h, N)$ consists of a document identifier $d \in \textsc{patos}$, a version identifier $v$, a cryptographic hash $h$ (SHA-256), and a set of PageIndex node identifiers $N = \{n_1, n_2, \ldots\}$ from which the information was extracted. The constraint $|P| \geq 1$ ensures that no Evidence Pack exists without at least one anchoring source.

    \item $\mathbf{L}$ (Epistemic Limits) is a 4-tuple $(\Delta, G, D, S)$ where:
    \begin{itemize}
        \item $\Delta$ (Divergences) is a set of identified disagreements between sources on the same topic;
        \item $G$ (Gaps) is a set of documentary gaps (information not available or not located);
        \item $D$ (Dependencies) is a set of clinical context conditions that modulate assertion validity;
        \item $S$ (Silences) is a set of normative silences, i.e., questions on which regulatory sources do not explicitly pronounce.
    \end{itemize}

    \item $\mathbf{C}$ (Curatorial Status) is a triple $(\sigma, c, j)$ where $\sigma \in \{\textit{draft}, \textit{under\_review}, \textit{accepted}, \textit{rejected}\}$ is the curation state, $c$ is the curator identifier, and $j$ is a natural-language justification for the curatorial decision.
\end{itemize}

\subsection{Well-formedness conditions}

An Evidence Pack $\mathbf{EP} = \langle Q, R, P, L, C \rangle$ is \textbf{well-formed} if and only if:

\begin{enumerate}
    \item \textbf{Typed question}: $\tau \in T$ (the question type belongs to the controlled taxonomy).
    \item \textbf{Source anchoring}: $|P| \geq 1$ (at least one provenance record exists).
    \item \textbf{Integrity verifiable}: for every $p_i = (d, v, h, N) \in P$, the hash $h$ matches the document $d$ at version $v$ in PATOS.
    \item \textbf{Node traceability}: for every $p_i$, the set $N$ is non-empty, i.e., the assertion is traceable to at least one specific document node.
    \item \textbf{Epistemic completeness}: $L$ is explicitly specified; an empty set in any component of $L$ is semantically distinct from the absence of $L$ (the former states ``no divergences/gaps/dependencies/silences were identified''; the latter is a specification error).
    \item \textbf{Curatorial closure}: if $\sigma \in \{\textit{accepted}, \textit{rejected}\}$, then $c \neq \emptyset$ and $j \neq \emptyset$ (accepted or rejected packs must have an identified curator and a documented justification).
\end{enumerate}

\subsection{Illocutionary typing}

The assertion type $\tau$ determines the illocutionary force of the Evidence Pack, following the speech act classification of Searle \cite{Searle1979}:

\begin{table}[ht]
\centering
\caption{Illocutionary typing of Evidence Pack assertion types, following Searle's speech act classification.}
\label{tab:illocutionary}
\small
\begin{tabular}{@{}lll@{}}
\toprule
\textbf{Assertion Type $\tau$} & \textbf{Illocutionary Class} & \textbf{Force} \\
\midrule
\textsc{indication} & Assertive & Commits to truth of therapeutic applicability \\
\textsc{contraindication} & Directive & Instructs the professional to avoid \\
\textsc{dosing} & Directive + Assertive & Prescribes conduct with factual basis \\
\textsc{interaction} & Assertive + Directive & States a fact and warns about consequences \\
\textsc{adverse\_reaction} & Assertive & Reports observed or documented effects \\
\textsc{warning} & Directive & Urges caution under specified conditions \\
\textsc{precaution} & Directive & Recommends monitoring or adjusted use \\
\textsc{special\_population} & Assertive + Directive & Qualifies applicability to a subgroup \\
\textsc{normative\_silence} & Non-commitment & Signals absence of regulatory pronouncement \\
\bottomrule
\end{tabular}
\end{table}

\noindent The type \textsc{normative\_silence} is epistemically distinguished: it does not assert, deny, permit, or prohibit, but formally registers that the normative system does not pronounce on the question posed. This aligns with the concept of \emph{normative gap} in deontic logic \cite{AlchourronBulygin1971,Navarro2014}.

\subsection{Lifecycle}

The Evidence Pack lifecycle follows a strictly forward progression:

\begin{lstlisting}
draft -> under_review -> accepted | rejected
\end{lstlisting}

\noindent State transitions are governed by:

\begin{itemize}
    \item \textbf{draft $\rightarrow$ under\_review}: triggered by completion of the Lector reading pipeline.
    \item \textbf{under\_review $\rightarrow$ accepted}: requires $c \neq \emptyset$, $j \neq \emptyset$, and explicit human validation.
    \item \textbf{under\_review $\rightarrow$ rejected}: requires $c \neq \emptyset$, $j \neq \emptyset$, with documented reason for rejection.
    \item \textbf{No backward transitions}: a rejected pack cannot be re-accepted; a new Evidence Pack must be created with a new provenance chain. This ensures auditability of the curatorial trajectory.
\end{itemize}

\subsection{Composition and derivation}

Evidence Packs are \textbf{compositional}: a complex pharmaceutical assertion (e.g., ``dipyrone is indicated for mild to moderate pain in adults, except in patients with known hypersensitivity or a history of agranulocytosis'') can be decomposed into multiple Evidence Packs, each with its own type, provenance chain, and epistemic limits.

Evidence Packs are \textbf{derivable but not transformable}: a new Evidence Pack may be created by consulting the same or different sources, but an existing accepted Evidence Pack is never modified; it is versioned. This mirrors the immutability principle of the PATOS layer.

\end{document}